\documentclass[notitlepage,nofootinbib]{revtex4-1}

\usepackage{graphicx}
\usepackage[justification=justified, singlelinecheck=off, compatibility=false]{caption}
\usepackage{subcaption}
\usepackage{amsmath}
\usepackage{bm}
\usepackage{amssymb}
\usepackage{color}
\usepackage{times}
\usepackage{float}
\usepackage{hyperref}
\usepackage[utf8]{inputenc}
\hypersetup{
    colorlinks=true,       
    linkcolor=red,          
    citecolor=blue,        
    filecolor=magenta,      
    urlcolor=blue           
}

\newcommand{\eref}[1]{Eq.~(\ref{#1})}
\newcommand{\fref}[1]{Fig.~\ref{#1}}
\newcommand{\sref}[1]{Sec.~\ref{#1}}
\newcommand{\tref}[1]{Table~\ref{#1}}


\begin{document}

\title{Biprobability approach to CP phase degeneracy from non-standard neutrino interactions}

\date{\today}

\author{Jeffrey M. Hyde}
\email{jeffrey.hyde@goucher.edu}
\affiliation{Department of Physics \& Astronomy, Goucher College, Baltimore, Maryland 21204, USA\footnote{Address beginning August 2019: Department of Physics and Astronomy, Bowdoin College, Brunswick, Maine 04011 USA}}

\begin{abstract}
Non-standard interactions (NSI) between neutrinos and matter at long-baseline experiments could make determination of the CP-violating phase $\delta_{13}$ ambiguous due to interference with additional complex phases. Such degeneracies are often studied in the context of specific experiments and a few parameter choices, leaving it unclear how to extract a general understanding of when two sets of parameters may be degenerate or how different types of experiments in principle combine to lift such a degeneracy. This work complements detailed simulations of individual experiments by showing how underlying parameters relate to degeneracies as represented on a biprobability plot. We show how a range of energies near the oscillation maximum $\Delta_{31} = \pi/2$ separates some degenerate probabilities along the CP-conserving direction of biprobability space according to $\delta_{+} \equiv \delta_{13} + \delta_{e\tau}$, while near $\Delta_{31} = 3\pi/2$ degenerate probabilities are separated along the CP-violating direction according to $\delta_{e\tau}$. We apply this to the experimental hints that suggest $\delta_{13} \sim -\pi/2$ to see that this could also be consistent with $\delta_{13}, \delta_{e\tau} = 0$ or $\pi$. The baseline and energy range characteristic of DUNE provides some resolution, but a further improvement comes from beams a few degrees off-axis at $\gtrsim 1000$ km baselines, including some proposed sites for T2HKK.

\vspace{1cm}
\end{abstract}

\maketitle

\setlength{\parindent}{20pt}

\section{Introduction}\label{sec:intro}

An important goal of the current and upcoming generation of long-baseline neutrino oscillation experiments is to measure the phase $\delta_{13}$ in the mixing matrix, which violates CP symmetry if its value is not 0 or $\pi$. Determining the existence and nature of leptonic CP violation is an important step in understanding the mixing properties in the standard 3-neutrino picture, and could have further implications for early universe leptogenesis \cite{Branco:2011zb,Hagedorn:2017wjy}. However, neutrinos could experience beyond-Standard Model interactions with the matter in between source and detector. This possibility is often studied in a model-independent way as an addition to the matter potential parametrized by a set of non-standard interaction (NSI) parameters \cite{Grossman:1995wx,Friedland:2004pp,Friedland:2004ah,Antusch:2008tz,Miranda:2015dra,Farzan:2015hkd}. These NSI parameters include new CP-violating phases that contribute to oscillation probabilities along with $\delta_{13}$, leading to potential ambiguity in determining the true underlying parameters.

It is possible, for instance, that for a given choice of underlying parameters and experimental setup (neutrino energy and baseline length), there could be significant CP violation present in the model but not apparent in the data. In other words, nature chooses parameters which then imply oscillation probabilities $P(\nu_{\mu} \rightarrow \nu_{e})$ and $P(\overline{\nu}_{\mu} \rightarrow \overline{\nu}_{e})$, while we measure events to obtain these probabilities which we hope will tell us about the parameters. But different underlying parameters can give the same or similar oscillation probabilities, leading to a parameter degeneracy.

Previous work has examined degeneracies that arise in the case of standard three-neutrino mixing without NSI \cite{Arafune:1997hd,Barger:2001yr,Minakata:2001qm,Ohlsson:2013ip}, for instance, normal versus inverted hierarchy and the octant of $\theta_{23}$. Degeneracies that arise due to NSI, especially CP phase degeneracies, have also been examined \cite{GonzalezGarcia:2001mp,Coloma:2011rq,Friedland:2012tq,Rahman:2015vqa,Masud:2015xva,Coloma:2015kiu,Palazzo:2015gja,deGouvea:2015ndi,Masud:2016bvp,Blennow:2016etl,Liao:2016hsa,deGouvea:2016pom,Liao:2016orc,Ge:2016dlx,Agarwalla:2016fkh,Fukasawa:2016lew,Deepthi:2016erc,deGouvea:2017yvn,Deepthi:2017gxg}. Such work often presents results in the experimentally well-motivated form of events as a function of neutrino energy, for instance in considering the potential of DUNE or T2HK to discover CP violation. The reach of DUNE and T2HK in resolving $\delta_{13}$ in the standard (no-NSI) case is examined in \cite{Ballett:2016daj}. Presently, experimental data does not rule out either mass ordering or any values of $\delta_{13}$, but there are experimental hints for CP nonconservation \cite{Abe:2017vif,NOvA:2018gge,Palazzo:2015gja}; we will return to this in \sref{sec:example}.

\begin{figure}[t]
	\centering
	\begin{subfigure}[b]{0.45\textwidth}
		\centering
		\includegraphics[width=\textwidth]{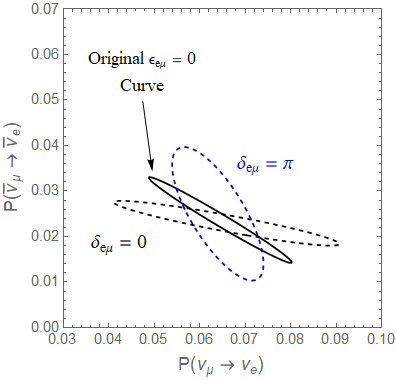}
		\caption{\label{fig:degeneracy_example}}
	\end{subfigure} \, \, \, \, \, \, %
	\begin{subfigure}[b]{0.45\textwidth}
		\centering
		\includegraphics[width=\textwidth]{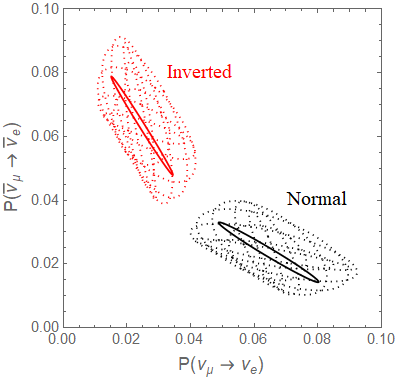}
		\caption{\label{fig:L1300_ellipse_fuzzball}}
	\end{subfigure}
\caption{\label{fig:ellipse_examples}Effect of nonzero NSI parameter $\epsilon_{e\mu}$ on biprobability plots ($\delta_{13}$ varies from 0 to $2\pi$ along each curve). The solid curves show the standard case with $\epsilon_{e\mu} = 0$, and the dashed curves show $|\epsilon_{e\mu}| = 0.05$ for several values of the CP violating phase $\delta_{e\mu}$. In \fref{fig:degeneracy_example} it is evident that a given $P$, $\overline{P}$ could correspond to widely separated values of $\delta_{e\mu}$, and even the no-NSI case can be degenerate with them. \fref{fig:L1300_ellipse_fuzzball} shows the range of possibilities by plotting dotted curves for values of $\delta_{e\mu}$ going from 0 to $2\pi$ in steps of $\pi/4$. Both plots are for $L = 1300$ km and correspond to $\Delta_{31} = \pi/2$, and \fref{fig:degeneracy_example} shows only the normal mass hierarchy. Parameters and numerical methods are discussed in \sref{sec:methods}.}
\end{figure}

Computing the spectrum of events as a function of neutrino energy is well-defined, but working backward to interpret the result in terms of underlying parameters can be unintuitive. Furthermore, results are often presented in terms of sensitivity plots for particular experimental setups, making it more difficult to discern what is a feature of the basic oscillation parameters and what is related to specifics of the experiment. While analytic expressions are available, it can be difficult to tell at a glance what may be the effect of changing a given parameter. Therefore, it's desirable to understand how to concisely organize the information in a way that promotes easy correspondence between oscillation probabilities and favored parameters or models. One method that has been useful for representing other degeneracies, and the situations where they are lifted, is to plot oscillation probability for antineutrinos, $\overline{P} \equiv {\rm Prob}(\overline{\nu}_{\mu}\rightarrow\overline{\nu}_{e})$, versus oscillation probability for neutrinos, $P \equiv {\rm Prob}(\nu_{\mu}\rightarrow\nu_{e})$, for values of $\delta_{13}$ ranging from $0 \rightarrow 2\pi$ \cite{Minakata:2001qm}. This gives a curve that traces out an ellipse in biprobability space, and is a useful way to visualize degeneracies for a range of parameter values. An example of such for neutrino oscillations with standard interactions is shown by the solid curve in \fref{fig:degeneracy_example}.

The biprobability plot also provides a visualization of parameter degeneracies and their complexity. This is apparent from the dashed curves in \fref{fig:degeneracy_example}, which show the result when $|\epsilon_{e\mu}|=0.05$, for a few values of $\delta_{e\mu}$.\footnote{Parameters and the values used in this paper are defined in \sref{sec:methods}. As we'll see later, the effect of $\epsilon_{e\mu}$ and $\epsilon_{e\tau}$ are qualitatively the same.} There are evidently some values of the neutrino and antineutrino probabilities that do not lead back to a unique choice of the underlying parameters. \fref{fig:L1300_ellipse_fuzzball} shows that this is a problem in general: the solid curves are for the normal and inverted mass orderings in the Standard Oscillation case (no NSI), and the dotted curves show biprobability ellipses for $|\epsilon_{e\mu}|=0.05$ and $\delta_{e\mu}$ varying from 0 to $2\pi$ in steps of $\pi/4$. The point emphasized by \fref{fig:ellipse_examples} but not unique to its parameter choices is that measuring $P$ and $\overline{P}$ is not generally enough to uniquely distinguish the underlying parameters, in particular to distinguish $\delta_{13}$ from $\delta_{e\mu}$ or $\delta_{e\tau}$.

Looking at different baseline lengths and/or neutrino energies helps resolve such degeneracies, and some previous work has incorporated biprobability plots into studies of NSI. The ability of a combination of 3000 km and 7000 km baseline experiments to distinguish CP violation due to NSI was considered in \cite{Gago:2009ij}. This work also showed how biprobability plots are affected by NSI, but did not quantify degeneracies in the manner that the present paper does. Biprobability plots have also been used to represent degeneracies in the presence of $\epsilon_{e\tau}$ at NO$\nu$A, with examples of degenerate probabilities at NO$\nu$A that are separated with the broader energy spectrum at DUNE \cite{Friedland:2012tq}, and to help represent an analytic result, showing that any ``apparent" $\delta_{13}$ and hierarchy in the standard case could correspond to $\delta_{13} = 0$ for some values of $\epsilon_{e\mu}$ and $\delta_{e\mu}$ (and similar for $\epsilon_{e\tau}$) \cite{Liao:2016hsa}. Implications of nonzero $\epsilon_{e\tau}$, the effect on biprobability plots, and the ability of DUNE to distinguish some values of $\delta_{13}$ and $\delta_{e\tau}$ was also considered in \cite{Soumya:2016enw,Agarwalla:2016fkh}. However, as mentioned above, it is still desirable to have a general picture of how oscillation parameters influence these degeneracies, separate from detailed experimental scenarios. Some work has used biprobability ellipses to represent $\epsilon_{e\tau}$ degeneracies relevant to an apparent $\delta_{13} \sim -\pi/2$ \cite{Flores:2018kwk}.

This paper shows how $\delta_{13}$ and NSI parameters influence the biprobability space representation of $\nu_{\mu} \rightarrow \nu_{\rm e}$ and $\overline{\nu}_{\mu} \rightarrow \overline{\nu}_{\rm e}$ oscillations. We apply this insight to study degeneracies between $\delta_{13}$ and NSI phases $\delta_{\rm e \mu}, \, \delta_{\rm e\tau}$. In the absence of NSI, there is an approximate degeneracy between values of $\delta_{13}$ with the same $\sin\delta_{13}$. In the biprobability context, looking at energies away from the oscillation maximum (at $\Delta_{31}$ in terminology to be introduced in \sref{sec:methods}) helps to break degeneracies, largely by increasing the separation between two points along the ``CP-conserving" (or $P+\overline{P}$) direction on the biprobability plot. This carries over to the case of nonzero NSI, where two points of the same $\sin\delta_{13}$ and the same $\delta_+ \equiv \delta_{13}+\delta_{e\mu}$ or $\delta_{13}+\delta_{e\tau}$ are approximately degenerate near the $\Delta_{31} = \pi/2$ maximum. As the energy varies from this maximum, $\delta_+$ (rather than $\delta_{13}$ or $\delta_{e\tau}$) plays the dominant role in separating points. On the other hand, near $\Delta_{31} = 3\pi/2$ it is $\delta_{e\tau}$ or $\delta_{e\mu}$ that is dominant, and this tends to separate degenerate points along the CP-violating (or $P^-$) direction. We examine the approximate degeneracy between $\delta_{13} = -\pi/2$ (i.e. $\delta_{13} = -\pi/2$ with no NSI, apparently maximal CP violation) and the CP-conserving situation $\delta_{13} = 0, \pi$, $\delta{e\tau} = 0, \pi$ for $|\epsilon_{e\tau}| = 0.02$. Using the above results, we see that examining the energy spectrum around $\Delta_{31} = \pi/2$ (relevant to DUNE) provides some separation between these points, while $\Delta_{31} \approx 3\pi/2$ (relevant to T2HKK) further improves upon this. The purpose of this paper is not to make detailed predictions about whether specific experiments will be able to resolve some parameters at a given statistical significance. In contrast, we seek to understand at a general level how the parameters relevant to the next decade of experiments work together to determine degeneracies and their breaking.

Before we examine degeneracies, \sref{sec:methods} will describe the numerical and analytic methods used in this paper. Then \sref{sec:biprob_degens} will apply these to understand degeneracies and how they are represented on the biprobability plot and \sref{sec:lifting_cp_degen} will show the role of the energy dependence in lifting the degeneracy in biprobability space. \sref{sec:example} applies these results to the interesting question of which degeneracies could be present if $\delta_{13}$ has apparently been measured as $3\pi/2$, and how this degeneracy breaking can be represented in biprobability space. \sref{sec:conclusion} summarizes the results and outlines future work that will build on the results presented here.

\section{Numerical and Analytic Computation of Oscillation Probabilities}\label{sec:methods}

In this section, we describe the specific methods and parameter choices used to compute oscillation probabilities in this paper. In particular, we consider NSI affecting propagation (i.e. matter potential) but not affecting interactions in the detector. In this case, NSI appear as parameters $\epsilon_{ij}$ in the flavor-basis interaction Hamiltonian\footnote{In particular, the NSI parameters $\epsilon_{\alpha\beta}$ relevant to propagation through matter are weighted sums of the NSI couplings of neutrinos to the various fermions, see \cite{Coloma:2015kiu}. We also do not consider NSI affecting neutrino production or detection.}:
\begin{align}\label{eq:nsi_ham}
	\mathcal{H}_{\rm int} & = \frac{1}{2E}\left( U \begin{pmatrix} \, \, 0 \, \, & \, & \, \\ \, & \Delta m^2_{21} & \, \\ \, & \, & \Delta m^2_{31} \end{pmatrix} U^{\dagger} + a \begin{pmatrix} 1 + \epsilon_{ee} & \epsilon_{e\mu} & \epsilon_{e\tau} \\ \left(\epsilon_{e\mu}\right)^{\ast} & \epsilon_{\mu\mu} & \epsilon_{\mu\tau} \\ \left(\epsilon_{e\tau}\right)^{\ast} & \left(\epsilon_{\mu\tau}\right)^{\ast} & \epsilon_{\tau\tau} \end{pmatrix} \right)
\end{align}
where the off-diagonal elements have complex phases defined by $\epsilon_{ij} \equiv |\epsilon_{ij}|\exp\left( i\delta_{ij} \right)$, the parameter encoding the matter potential is
\begin{align}
	a & = 2 \, \sqrt{2} \, G_{\rm F} \, N_{\rm e} \, E \, = \, \left( 7.56 \times 10^{-5} \, {\rm eV}^2 \right) \left( \frac{\rho}{\rm g / cm^{3}} \right) \left( \frac{E}{\rm GeV} \right)
\end{align}
and the mixing matrix in terms of $s_{ij} \equiv \sin\theta_{ij}$, $c_{ij} \equiv \cos\theta_{ij}$ is
\begin{align}
	U & = \begin{pmatrix} 1 & 0 & 0 \\ 0 & c_{23} & s_{23} \\ 0 & -s_{23} & c_{23} \end{pmatrix} \begin{pmatrix} c_{13} & 0 & s_{13}e^{-i\delta_{13}} \\ 0 & 1 & 0 \\ -s_{13}e^{i\delta_{13}} & 0 & c_{13} \end{pmatrix} \begin{pmatrix} c_{12} & s_{12} & 0 \\ -s_{12} & c_{12} & 0 \\ 0 & 0 & 1 \end{pmatrix}
\end{align}
where $L$ is the baseline length, $E$ is the neutrino energy, $N_e$ is the number density of electrons, $\rho$ is the density of matter that neutrinos propagate through, and $G_{\rm F}$ is the Fermi constant.  \tref{tab:parameters} gives the values of various neutrino oscillation parameters used for the calculations in this paper. These are based on the 2016 Particle Data Group (PDG) Review of Particle Physics \cite{Patrignani:2016xqp}. In particular, one dimensionless combination that will come up often in the context of perturbation expansion is\footnote{Here ``$L$[km]" means the dimensionless numerical value of baseline length expressed in km, and similarly with $\rho$. For the rest of this paper, we will use $L$ and $\rho$ in this same manner, as dimensionless numbers in expressions for probabilities.}
\begin{align}\label{eq:alovere}
	\frac{a L}{ (\hbar c) E} & = (3.83 \times 10^{-4} ) \, L[{\rm km}] \, \rho[{\rm g / cm}^{3}].
\end{align}
Since the purpose of this paper is to gain a general understanding of the behavior of biprobability plots within reasonable parameter ranges, not rigorous implementation experimental constraints or simulation of specific experiments or detectors, we examine a range of NSI parameters inspired by the different bounds listed in \tref{tab:parameters}. These are completely model-independent bounds (``NSI Range A") and more restrictive but model-dependent bounds (``NSI Range B") listed in section VI.A of \cite{Ohlsson:2012kf}. In this work, we will examine $\epsilon_{e\mu}, \epsilon_{e\tau} \sim \mathcal{O}(0.01)$ to $\mathcal{O}(0.1)$. The reason for the lower limit is that for small enough $\epsilon_{\alpha\beta}$, CP phase degeneracies will not be significant, a point we will return to in \sref{sec:biprob_degens}.

\begin{table}[h]
\centering
	\begin{tabular}{|c|c||c||c|}
	\hline \, \textbf{Parameter} \, & \, \textbf{Value} \, & \, \textbf{NSI Range A} \, & \, \textbf{NSI Range B} \, \\ \hline
	$\theta_{12}$ & 0.587 & $|\epsilon_{ee}| < 2.5$ & $-0.9 < \epsilon_{ee} < 0.75$ \\ \hline
	$\theta_{23}$ & 0.80 & \, $|\epsilon_{\mu\mu}| < 0.046$ \, & \, $-0.05 < \epsilon_{\mu\mu} < 0.08$ \, \\ \hline
	$\theta_{13}$ & 0.145 & $|\epsilon_{\tau\tau}| < 9.0$ & $|\epsilon_{\tau\tau}| \lesssim 0.4$ \\ \hline
	$\Delta m^2_{21}$ & \, $7.53 \times 10^{-5} \, {\rm eV}^2$ \, & $|\epsilon_{e\mu}| < 0.21$ & $|\epsilon_{e\mu}| \lesssim 3.8 \times 10^{-4}$ \\ \hline
	$\Delta m^2_{32}$ & \, $2.45 \times 10^{-3} \, {\rm eV}^2$ \, & $|\epsilon_{e\tau}| < 1.7$ & $|\epsilon_{e\tau}| \lesssim 0.25$ \\ \hline
	$\Delta m^2_{31}$ & \, $\Delta m^2_{21} \pm \Delta m^2_{32}$ \, $\left( {\rm NH} / {\rm IH} \right)$ \, & $|\epsilon_{\mu\tau}| < 0.21$ & $|\epsilon_{\mu\tau}| \lesssim 0.25$ \\ \hline
	\end{tabular}
\caption{\label{tab:parameters}The standard mixing parameters in the left column are based on the 2016 Particle Data Group (PDG) Review of Particle Physics \cite{Patrignani:2016xqp}. The NSI parameter ranges are based on model-independent (``NSI Range A") and more model-dependent (``NSI Range B") bounds as described in section VI.A of \cite{Ohlsson:2012kf}.}
\end{table}

In matter, $P$ and $\overline{P}$ may generally be found by numerical integration: the time-evolution (equivalently, distance-evolution) of amplitudes follows from the Hamiltonian. Oscillation probabilities shown in this paper are the result of such a numerical integration using the Hamiltonian \eref{eq:nsi_ham}, with matter density $\rho = 3.0$ g/cm$^3$. Realistic variations in $\rho$ will change the results quantitatively but not qualitatively. We will restrict our attention to the electron neutrino (antineutrino) appearance  processes $\nu_{\mu} \rightarrow \nu_{\rm e}$ ($\overline{\nu}_{\mu} \rightarrow \overline{\nu}_{\rm e}$) studied at T2K, NO$\nu$A, and DUNE. These experiments aim for a combination of baseline length and neutrino energy giving $\Delta_{31} \equiv \Delta m^2_{31} L / (4 E ) \sim \pm 0.003 \, L$[km] / $E$[GeV] $= \pi / 2$, near the oscillation maximum. Results in this paper use $L = 1300$ km when necessary in order to give a sense of the size of effect that may be relevant to DUNE, but this should not be interpreted as a precise statement of what will be measured at the DUNE experiment. Relating our results to the capabilities of specific experiments is left to future work.

As an alternative to the numerical approach, it is often convenient to treat the parameters $\Delta m^2_{21} / \Delta m^2_{31}$, $\sin(\theta_{13})$, and $|\epsilon_{\alpha\beta}|$ as small parameters of the same order \cite{Arafune:1997hd,Cervera:2000kp,Kikuchi:2008vq}. Then oscillation probabilities may be represented by approximate analytic solutions. While not exact, this approach is still helpful in illuminating trends seen in numerical results, as we will see in the remaining sections. Before we get to the full perturbative expressions, we will point out some important general features.

In vacuum, the Standard Model oscillation probabilities can be expressed in the form
\begin{align}\label{eq:ellipse_form}
{\rm Prob}(\nu_{\mu}\rightarrow\nu_e) \, \equiv P & = N + C_{13}\cos\delta_{13} + S_{13}\sin\delta_{13}, \nonumber \\
{\rm Prob}(\overline{\nu}_{\mu}\rightarrow\overline{\nu}_e) \, \equiv \overline{P} & = \overline{N} + \overline{C}_{13}\cos\delta_{13} + \overline{S}_{13}\sin\delta_{13},
\end{align}
which is the parametric form for an ellipse in the $P, \overline{P}$ plane as $\delta_{13}$ varies from 0 to $2\pi$ \cite{Minakata:2001qm}. Using the perturbation expansion \cite{Cervera:2000kp} in matter, the oscillation probabilities still take the form of \eref{eq:ellipse_form}, but with the coefficients $N$, $C_{13}$, etc. altered by the matter effects. In the perturbation expansion with NSI present, the form of \eref{eq:ellipse_form} remains \cite{Kikuchi:2008vq}, i.e. varying $\delta_{13}$ still traces an ellipse in biprobability space. For later use in this paper, we will refer to the standard oscillation results as $P_0, \overline{P}_0$ and the effect of NSI as a $\Delta P_{e\mu}, \Delta \overline{P}_{e\mu}$ or $\Delta P_{e\tau}, \Delta \overline{P}_{e\tau}$ that would be added to $P_0, \overline{P}_0$:
\begin{align}\label{eq:general_deltap_defn}
	P & = P_0 + \Delta P_{e\mu}, \, \, \, \, \, \, \, \, \, \, \overline{P} \, = \, \overline{P}_0 + \Delta \overline{P}_{e\mu}
\end{align}
and the same expression for $\epsilon_{e\tau}$ with $\Delta P_{e\mu} \rightarrow \Delta P_{e\tau}$. It's interesting to note that the probabilities can also be expressed in a form that factors out sines and cosines of the ``hidden sector" phase $\delta_{e\mu}$ rather than $\delta_{13}$:
\begin{align}\label{eq:nsiemu_ppbar_hsellipse}
	P & = P_0 + \Delta S_{e\mu}\sin\delta_{e\mu} + \Delta C_{e\mu}\cos\delta_{e\mu} \nonumber \\
	\overline{P} & = \overline{P}_0 + \Delta \overline{S}_{e\mu}\sin\delta_{e\mu} + \Delta \overline{C}_{e\mu}\cos\delta_{e\mu}
\end{align}
and the same can be done for $\delta_{e\tau}$. While the form of \eref{eq:ellipse_form} is an ellipse along which $\delta_{13}$ continuously varies from 0 to $2\pi$, \eref{eq:nsiemu_ppbar_hsellipse} describes a biprobability ellipse along which $\delta_{e\mu}$ continuously varies for one constant value of $\delta_{13}$. Examples of these ``hidden sector" ellipses from numerical integration are shown in \fref{fig:L1300_demu_ellipses_nh}, superposed on the standard biprobability ellipse that varies $\delta_{13}$ when $\epsilon_{e\mu} = 0$. Previous works have also used this form \cite{Gago:2009ij,Soumya:2016enw,Agarwalla:2016fkh,Forero:2016cmb}, and in \sref{sec:biprob_degens} we will find it particularly useful for representing degeneracies.

In the standard oscillation case, the perturbative expansion \cite{Arafune:1997hd,Cervera:2000kp,Kikuchi:2008vq} gives probabilities
\begin{align}\label{eq:prob_zero}
	P_0, \, \overline{P}_0 & = \left[ \sin^2(2\theta_{13})s_{23}^2\sin^2(\Delta_{31}) + c_{23}^2\sin^2(2\theta_{12})\left( \frac{\Delta m^2_{21}}{\Delta m^2_{31}} \right)^2\Delta_{31}^2 \right. \nonumber \\
	& \left. \, \, \, \, \, \, \, \, \, \, \, \, \, \, \, \, \, \, \, \, \, \, \, \, \, \, \, \, \, \, \, \, \, \, \, \, \, \, \, \, \, \, \, \, \, \, \, \, \, \, \, \, \, \, \, \, \, \, \, \, \, \, \, \, \, \, \, \, \, \, \, \, \, \, \, \, \, \, \, \, \, \, \, \, \, \, \, \, \, \, \, \, \, \, \, \, \, \, \, \, \, \, \, \, \, \, \, \, \, \, \, \, \, \, \, \, \, \, \, \, \, \, \, \, \, \, \, \, \, \, \, \, \, \, \, + 4 J_r \left( \frac{\Delta m^2_{21}}{\Delta m^2_{31}} \right)\Delta_{31}\sin(2\Delta_{31})\cos(\delta_{13}) \right] \nonumber \\
	& \, \, \, \, \, \pm \left[ -8J_r\left( \frac{\Delta m^2_{21}}{\Delta m^2_{31}} \right)\Delta_{31}\sin^2(\Delta_{31})\sin(\delta_{13}) + \frac{aL}{2E}s_{23}^2\sin^2(2\theta_{13})\left( \frac{\sin^2(\Delta{31})}{\Delta_{31}} - \frac{1}{2}\sin(2\Delta_{31}) \right) \right],
\end{align}
where $+$ ($-$) corresponds to $P_0$ ($\overline{P}_0$) and $J_r \equiv c_{12}s_{12}c_{13}^2s_{13}c_{23}s_{23}$ (with $c_{12} \equiv \cos(\theta_{12})$ etc.). With $\epsilon_{e\mu} \neq 0$ but all other NSI parameters absent \cite{Kikuchi:2008vq}, the probabilities in \eref{eq:prob_zero} are adjusted by
\begin{align}\label{eq:deltaprob_emu}
	\Delta P_{e\mu}, \, \Delta\overline{P}_{e\mu} & = \frac{2aL}{E}|\epsilon_{e\mu}|\left[ -s_{13}s_{23}c_{23}^2\sin^2(\Delta_{31})\sin(\delta_+) + c_{12}s_{12}c_{23}s_{23}^2 \left( \frac{\Delta m^2_{21}}{\Delta m^2_{31}} \right) \sin^2(\Delta_{31})\sin(\delta_{e\mu}) \right] \nonumber \\
	& \, \, \, \, \, \pm \frac{2aL}{E}|\epsilon_{e\mu}| \left[ s_{13}s_{23}\left( s_{23}^2\frac{\sin^2(\Delta_{31})}{\Delta_{31}} - \frac{1}{2}c_{23}^2\sin(2\Delta_{31}) \right)\cos(\delta_+) \right. \nonumber \\
	& \, \, \, \, \, \, \, \, \, \, \, \, \, \, \, \, \, \, \, \, \, \, \, \, \, \, \, \, \, \, \, \, \, \, \, \, \, \, \, \, \, \, \, \, \, \left. + c_{12}s_{12}c_{23}\left( \frac{\Delta m^2_{21}}{\Delta m^2_{31}} \right) \left( c_{23}^2\Delta_{31} + \frac{1}{2}s_{23}^2\sin(2\Delta_{31}) \right) \cos(\delta_{e\mu})\right].
\end{align}
With $\epsilon_{e\tau} \neq 0$ but all other NSI parameters absent \cite{Kikuchi:2008vq}, the probabilities are adjusted by
\begin{align}\label{eq:deltaprob_etau}
	\Delta P_{e\tau}, \, \Delta\overline{P}_{e\tau} & = \frac{2aL}{E}|\epsilon_{e\tau}|\left[ s_{13}c_{23}s_{23}^2\sin^2(\Delta_{31})\sin(\delta_+) + c_{12}s_{12}s_{23}c_{23}^2 \left( \frac{\Delta m^2_{21}}{\Delta m^2_{31}} \right) \sin^2(\Delta_{31})\sin(\delta_{e\tau}) \right] \nonumber \\
	& \, \, \, \, \, \pm \frac{2aL}{E}|\epsilon_{e\tau}| \left[ s_{13}c_{23}s_{23}^2\left( \frac{\sin^2(\Delta_{31})}{\Delta_{31}} - \frac{1}{2}\sin(2\Delta_{31}) \right)\cos(\delta_+) \right. \nonumber \\
	& \, \, \, \, \, \, \, \, \, \, \, \, \, \, \, \, \, \, \, \, \, \, \, \, \, \, \, \, \, \, \, \, \, \, \, \, \, \, \, \, \, \, \, \, \, \left. - c_{12}s_{12}s_{23}c_{23}^2\left( \frac{\Delta m^2_{21}}{\Delta m^2_{31}} \right) \left( \Delta_{31} - \frac{1}{2}\sin(2\Delta_{31}) \right) \cos(\delta_{e\tau}) \right],
\end{align}
where
\begin{align}\label{eq:delta_plus_def}
	\delta_+ & \equiv \delta_{13} + \delta_{e\mu} \, \, \, \, \, {\rm or} \, \, \, \, \, \delta_{13} + \delta_{e\tau}
\end{align}
To keep the notation simple, we use $\delta_+$ in the case of either nonzero $\epsilon_{e\mu}$ or $\epsilon_{e\tau}$. In this paper we will consider either one at a time to be nonzero, so from context the definition of $\delta_+$ will be clear.

\section{CP Phase Degeneracies in the Presence of NSI}\label{sec:biprob_degens}

We will now use the methods reviewed in \sref{sec:methods} to develop an understanding of degeneracies in the presence of NSI. In \sref{sec:formalism} we will put the perturbative expressions into a more convenient form for the examination of CP Violation in biprobability space, and in \sref{sec:degens} we will use this along with numerical solution of probabilities in order to study degeneracies. Ref. \cite{Agarwalla:2016fkh} also used this combination of methods to study the effect of NSI as seen on biprobability plots and agrees with our results when there is overlap, but here we have a narrower scope and focus more on quantifying phase degeneracies and their breaking.

\subsection{Applying the Perturbative Expressions}\label{sec:formalism}

It will be convenient to note that $P$ takes the schematic form of
\begin{align}\label{eq:cp_even_odd}
	P & = \left( {\rm CP \, even \, terms} \right) + \left( {\rm CP \, odd \, terms} \right), \nonumber \\
	{\rm so} \, \, \, \, \, \, \, \, \, \, \overline{P} & = \left( {\rm CP \, even \, terms} \right) - \left( {\rm CP \, odd \, terms} \right).
\end{align}
While previous works (e.g. \cite{Masud:2016bvp}) have considered such a decomposition, here we will take this idea further and see that it leads to a useful result. In biprobability space, \eref{eq:cp_even_odd} suggests that the rotated coordinates
\begin{align}\label{eq:pplus_pminus_def}
	P^+ & \equiv \frac{1}{\sqrt{2}}(P + \overline{P}), \, \, \, \, \, P^- \, \equiv \, \frac{1}{\sqrt{2}}(P - \overline{P})
\end{align}
are natural to look at in the context of CP violation, since
\begin{align}
	P^+ & = \sqrt{2} \left( {\rm CP \, even \, terms} \right), \, \, \, \, \, P^- \, = \, \sqrt{2} \left( {\rm CP \, odd \, terms} \right).
\end{align}
We can see this explicitly using the perturbative expressions: \eref{eq:prob_zero} gives
\begin{align}\label{eq:prob_zero_plusminus}
	P_0^+ & = \sqrt{2}\sin^2(2\theta_{13})s_{23}^2\sin^2(\Delta_{31}) + \sqrt{2}c_{23}^2\sin^2(2\theta_{12})\left( \frac{\Delta m^2_{21}}{\Delta m^2_{31}} \right)^2\Delta_{31}^2 \nonumber \\
	& \, \, \, \, \, \, \, \, \, \, \, \, \, \, \, \, \, \, \, \, \, \, \, \, \, \, \, \, \, \, \, \, \, \, \, \, \, \, \, \, \, \, \, \, \, \, \, \, \, \, \, \, \, \, \, \, \, \, \, \, \, \, \, \, \, \, \, \, \, \, \, \, \, \, \, \, \, \, \, \, \, \, \, \, \, \, \, \, \, \, \, \, \, \, \, \, \, \, \, \, \, \, \, \, \, \, \, \, \, \, \, \, \, \, \, \, \, \, \, \, + 4\sqrt{2} J_r \left( \frac{\Delta m^2_{21}}{\Delta m^2_{31}} \right)\Delta_{31}\sin(2\Delta_{31})\cos(\delta_{13}) \nonumber \\
	P_0^- & = -8\sqrt{2}J_r\left( \frac{\Delta m^2_{21}}{\Delta m^2_{31}} \right)\Delta_{31}\sin^2(\Delta_{31})\sin(\delta_{13}) + \frac{aL}{\sqrt{2}E}s_{23}^2\sin^2(2\theta_{13})\left( \frac{\sin^2(\Delta_{31})}{\Delta_{31}} - \frac{1}{2}\sin(2\Delta_{31}) \right),
\end{align}
\eref{eq:deltaprob_emu} gives
\begin{align}\label{eq:deltaprob_emu_plusminus}
	\Delta P_{e\mu}^+ & = \frac{2\sqrt{2}aL}{E}s_{23}c_{23}\sin^2(\Delta_{31})|\epsilon_{e\mu}|\left[ -s_{13}c_{23}\sin(\delta_+) + c_{12}s_{12}s_{23} \left( \frac{\Delta m^2_{21}}{\Delta m^2_{31}} \right) \sin(\delta_{e\mu}) \right], \nonumber \\
	\Delta P_{e\mu}^- & = \frac{2\sqrt{2}aL}{E}s_{23}^2|\epsilon_{e\mu}| \left[ s_{13}s_{23}\left( \frac{\sin^2(\Delta_{31})}{\Delta_{31}} - \frac{1}{2}\left(\frac{c_{23}}{s_{23}}\right)^2\sin(2\Delta_{31}) \right)\cos(\delta_+) \right. \nonumber \\
	& \, \, \, \, \, \, \, \, \, \, \, \, \, \, \, \, \, \, \, \, \, \, \, \, \, \, \, \, \, \, \, \, \, \, \, \, \, \, \, \, \, \, \, \, \, \, \, \, \, \, \, \, \, \, \, \left. + c_{12}s_{12}c_{23}\left( \frac{\Delta m^2_{21}}{\Delta m^2_{31}} \right) \left( \left(\frac{c_{23}}{s_{23}}\right)^2\Delta_{31} + \frac{1}{2}\sin(2\Delta_{31}) \right) \cos(\delta_{e\mu})\right],
\end{align}
and \eref{eq:deltaprob_etau} gives
\begin{align}\label{eq:deltaprob_etau_plusminus}
	\Delta P_{e\tau}^+ & = \frac{2\sqrt{2}aL}{E}c_{23}s_{23}\sin^2(\Delta_{31})|\epsilon_{e\tau}|\left[ s_{13}s_{23}\sin(\delta_+) + c_{12}s_{12}c_{23} \left( \frac{\Delta m^2_{21}}{\Delta m^2_{31}} \right) \sin(\delta_{e\tau}) \right], \nonumber \\
	\Delta P_{e\tau}^- & = \frac{2\sqrt{2}aL}{E}c_{23}s_{23}|\epsilon_{e\tau}| \left[ s_{13}s_{23}\left( \frac{\sin^2(\Delta_{31})}{\Delta_{31}} - \frac{1}{2}\sin(2\Delta_{31}) \right)\cos(\delta_+) \right. \nonumber \\
	& \, \, \, \, \, \, \, \, \, \, \, \, \, \, \, \, \, \, \, \, \, \, \, \, \, \, \, \, \, \, \, \, \, \, \, \, \, \, \, \, \, \, \, \, \, \, \, \, \, \, \, \, \, \, \, \, \, \, \, \, \left. - c_{12}s_{12}c_{23}\left( \frac{\Delta m^2_{21}}{\Delta m^2_{31}} \right) \left( \Delta_{31} - \frac{1}{2}\sin(2\Delta_{31}) \right) \cos(\delta_{e\tau}) \right].
\end{align}
Using the known values of some parameters as outlined in \tref{tab:parameters}, \eref{eq:prob_zero_plusminus} becomes
\begin{align}\label{eq:prob_zero_plusminus_num}
	P_0^+ & = 0.059\sin^2(\Delta_{31}) + 0.00052\Delta_{31}^2 + 0.0055\Delta_{31}\sin(2\Delta_{31})\cos(\delta_{13}) \nonumber \\
	P_0^- & = -0.011\Delta_{31}\sin^2(\Delta_{31})\sin(\delta_{13}) + \left( 1.1\times 10^{-5} \right) \, L \, \rho \, \left( \frac{\sin^2(\Delta_{31})}{\Delta_{31}} - \frac{1}{2}\sin(2\Delta_{31}) \right),
\end{align}
\eref{eq:deltaprob_emu_plusminus} becomes 
\begin{align}\label{eq:deltaprob_emu_plusminus_num}
	\Delta P_{e\mu}^+ & = \left( 5.4 \times 10^{-5} \right) \, L \, \rho \, \sin^2(\Delta_{31}) \, |\epsilon_{e\mu}| \left[ -\sin(\delta_+) + 0.099 \sin(\delta_{e\mu}) \right], \nonumber \\
	\Delta P_{e\mu}^- & = \left( 5.6\times 10^{-5} \right) \, L \, \rho \, |\epsilon_{e\mu}| \left[ \left( \frac{\sin^2(\Delta_{31})}{\Delta_{31}} - 0.47\sin(2\Delta_{31}) \right)\cos(\delta_+) \right. \nonumber \\
	& \, \, \, \, \, \, \, \, \, \, \, \, \, \, \, \, \, \, \, \, \, \, \, \, \, \, \, \, \, \, \, \, \, \, \, \, \, \, \, \, \, \, \, \, \, \, \, \, \, \, \, \, \, \, \, \, \, \, \, \, \, \, \, \, \, \, \, \, \, \, \, \left. + 0.048\left(1.9\Delta_{31} + \sin(2\Delta_{31}) \right) \cos(\delta_{e\mu}) \right],
\end{align}
and \eref{eq:deltaprob_etau_plusminus} becomes
\begin{align}\label{eq:deltaprob_etau_plusminus_num}
	\Delta P_{e\tau}^+ & = \left( 5.4\times 10^{-5} \right) \, L \, \rho \, \sin^2(\Delta_{31}) \, |\epsilon_{e\tau}|\left[ \sin(\delta_+) + 0.096 \sin(\delta_{e\tau}) \right], \nonumber \\
	\Delta P_{e\tau}^- & = \left( 5.4 \times 10^{-5} \right) \, L \, \rho \, |\epsilon_{e\tau}| \left[ \left( \frac{\sin^2(\Delta_{31})}{\Delta_{31}} - \frac{1}{2}\sin(2\Delta_{31}) \right)\cos(\delta_+)  \right. \nonumber \\
	& \, \, \, \, \, \, \, \, \, \, \, \, \, \, \, \, \, \, \, \, \, \, \, \, \, \, \, \, \, \, \, \, \, \, \, \, \, \, \, \, \, \, \, \, \, \, \, \, \, \, \, \, \, \, \, \, \, \, \, \, \, \, \, \, \, \, \, \, \, \, \, \left. - 0.096 \left( \Delta_{31} - \frac{1}{2}\sin(2\Delta_{31}) \right) \cos(\delta_{e\tau}) \right].
\end{align}
We focus on long-baseline experiments with $\Delta_{31} \sim \pi/2$ or $3\pi/2$, so the $\Delta_{31}$ and $1/\Delta_{31}$ factors above will be $\sim \mathcal{O}(1)$ while $\sin(2\Delta_{31})$ will be equal or close to zero. \sref{sec:lifting_cp_degen} will take a closer look at the effect of the $\Delta_{31}$-dependent terms as the energy varies.

\subsection{Representing and Quantifying Degeneracies}\label{sec:degens}

We are now in a better position to quantitatively study degeneracies using the results of \sref{sec:formalism} and the ``hidden sector" biprobability ellipses that vary $\delta_{e\mu}$ or $\delta_{e\tau}$ for some fixed value of $\delta_{13}$. In this section, we will see that these are useful because to a good approximation, at $\Delta_{31} \approx \pi/2$ the ``center" is determined by $\sin(\delta_{13})$ and the dimensions (i.e. major and minor axes) are determined by $\epsilon_{e\mu}$ or $\epsilon_{e\tau}$. In other words, hidden sector ellipses with the same magnitude NSI parameter and same $\delta_{13}$ or $\pi - \delta_{13}$ are approximately degenerate, and a given point on the ellipse only determines $\delta_+$. Previous works (see e.g. \cite{Friedland:2012tq}) have used an approximation that helps give a sense of when points are degenerate: points separated by $\lesssim 0.01$ on the biprobability plot won't be resolved. While this does not fully represent a particular experiment's ability to resolve parameters, our aim here is to study at a general level when degeneracies may exist or be broken, so this convenient rule of thumb is sufficient for our purposes. In \sref{sec:lifting_cp_degen}, we will see that this degeneracy no longer holds for an energy-baseline combination away from $\Delta_{31} \approx \pi/2$.

From \eref{eq:prob_zero_plusminus_num} it is apparent that, in the absence of NSI, the probabilities in the ``CP-violating direction" $P_0^-$ are controlled by both $\sin\delta_{13}$ and matter effects, while probabilities in the ``non-CP-violating direction" are controlled by $\cos\delta_{13}$. Writing the probabilities in this form, we can see how matter effects help break the degeneracy between normal and inverted mass ordering. Under NH $\leftrightarrow$ IH, $\Delta_{31}$ changes sign while the magnitude is only slightly affected. Since both terms multiplying $aL/E$ change sign under $\Delta_{31} \rightarrow -\Delta_{31}$ ($\Delta_{31}$ and $\sin(2\Delta_{31})$ change sign, while $\sin^2(\Delta_{31})$ does not), the effect at first order is just to reflect the ellipse across $P^- = 0$. This role of the matter effect was noted in \cite{Minakata:2001qm} and is evident in the solid curves (standard oscillation case) in \fref{fig:L1300_ellipse_fuzzball} of this paper.

\begin{figure}[t]
	\centering
	\begin{subfigure}[b]{0.45\textwidth}
		\centering
		\includegraphics[width=\textwidth]{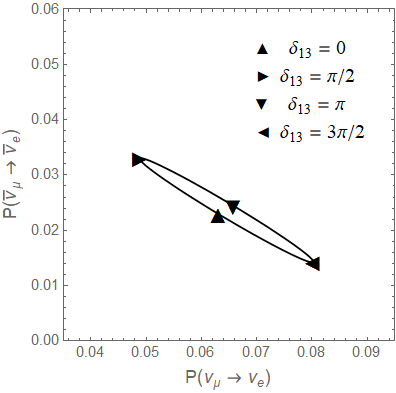}
		\caption{\label{fig:L1300_x0_SM_NH}}
	\end{subfigure} \, \, \, \, \, \, %
	\begin{subfigure}[b]{0.45\textwidth}
		\centering
		\includegraphics[width=\textwidth]{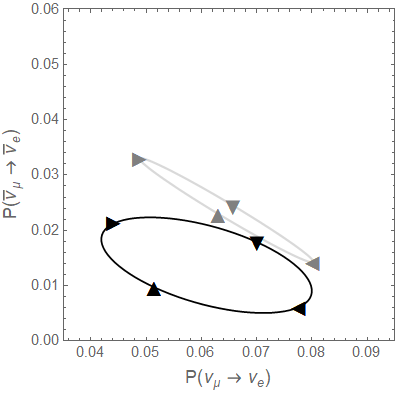}
		\caption{\label{fig:L1300_xm0p2_SM_NH}}
	\end{subfigure}
\caption{\label{fig:sm_nh_degens}Biprobability ellipse for $L=1300$ km and $\epsilon_{e\mu} = \epsilon_{e\tau} = 0$ as $\delta_{13}$ varies continuously from $0$ to $2\pi$. A few values of $\delta_{13}$ are marked (the triangular markers can be thought of as a clock hand going around once every $2\pi$). \fref{fig:L1300_x0_SM_NH} corresponds to $\Delta_{31} = \pi/2$. The degeneracies mentioned in \eref{eq:d13_degen} are evident. \fref{fig:L1300_xm0p2_SM_NH} corresponds to $E=2.0$ GeV, so $\Delta_{31}$ no longer equals $\pi/2$ (and $\sin(2\Delta_{31})\neq 0$), and the degeneracy lifts. (The gray lines show the same values plotted in \fref{fig:L1300_x0_SM_NH} for ease of comparison.)}
\end{figure}

Another insight from the case of standard interactions relates to $\delta_{13}$. Taking as an example $L=1300$ km and $\rho=3$ g/cm$^3$, \eref{eq:prob_zero_plusminus_num} gives
\begin{align}\label{eq:d13_degen}
	P_0^+ &= 0.060, \, \, \, \, \, \, \, \, \, \, P_0^- \, = \, -0.017\sin(\delta_{13}) + 0.027.
\end{align}
There is evidently a degeneracy between a given $\delta_{13}$ and $\pi - \delta_{13}$, since the sine of each is the same. \fref{fig:L1300_x0_SM_NH} shows this approximate degeneracy (note that the approximate result giving \eref{eq:prob_zero_plusminus_num} does not exactly predict $P$ and $\overline{P}$, as they are not exactly degenerate, but it does accurately describe the general behavior). Now considering nonzero $\epsilon_{e\mu}$ or $\epsilon_{e\tau}$, in \eref{eq:deltaprob_emu_plusminus_num} the coefficient of the $\delta_+$ term is appreciably greater than the coefficient of the $\delta_{e\mu}$ term, for both $\Delta P_{e\mu}^+$ and $\Delta P_{e\mu}^-$. The same is true in \eref{eq:deltaprob_etau_plusminus_num}. This suggests that hidden sector ellipses (constant $\delta_{13}$, $\delta_{e\mu}$ varying from 0 to $2\pi$) with $\delta_{13}$ and $\pi - \delta_{13}$ should approximately overlap.

\begin{figure}[t]
	\centering
	\begin{subfigure}[b]{0.45\textwidth}
		\centering
		\includegraphics[width=\textwidth]{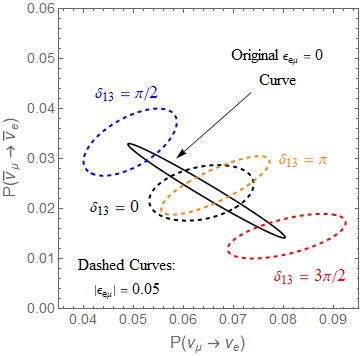}
		\caption{\label{fig:L1300_demu_ellipses_nh}}
	\end{subfigure} \, \, \, \, \, \, %
	\begin{subfigure}[b]{0.45\textwidth}
		\centering
		\includegraphics[width=\textwidth]{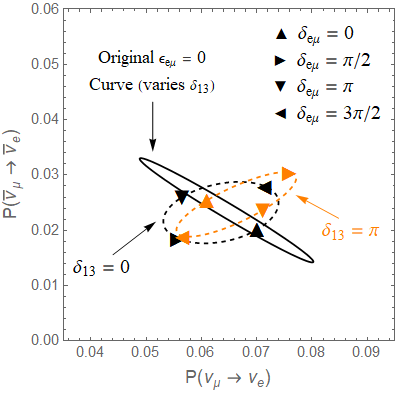}
		\caption{\label{fig:overlapping_ellipses_nh_deltaminus}}
	\end{subfigure}
\caption{\label{fig:demu_ellipse_variation}Example of biprobability plots where solid curves correspond to the standard ellipse that varies $\delta_{13}$, with $|\epsilon_{e\mu}|=0$ and dashed curves correspond to the ``hidden sector ellipse" that varies $\delta_{e\mu}$ with fixed $\delta_{13}$, as discussed after \eref{eq:nsiemu_ppbar_hsellipse}. \fref{fig:L1300_demu_ellipses_nh} shows (for NH) hidden sector ellipses for four values of $\delta_{13}$; these can be thought of as ``how a given point on the biprobability curve changes due to NSI." In particular, there is an approximate degeneracy between the $\delta_{13} = 0$ and $\delta_{13} = \pi$ curves, which \fref{fig:overlapping_ellipses_nh_deltaminus} focuses on, also giving specific values of $\delta_{e\mu}$ as discussed in the text.}
\end{figure}

Numerical results illustrate this, for instance by looking at $\delta_{13} = 0, \pi$. For energy and baseline length corresponding to $\Delta_{31} = \pi/2$, \eref{eq:deltaprob_emu_plusminus_num} gives the same effect $\Delta P_{e\mu}^+ \approx 0$, $\Delta P_{e\mu}^- \approx +0.007$ for two different points, $\delta_{13} = 0, \delta_{e\mu} = 0$ and $\delta_{13} = \pi, \delta_{e\mu} = \pi$, which both have $\delta_+ = 0$. On the other hand, for two points that have $\delta_+ = \pi$, namely $\delta_{13} = 0, \, \delta_{e\mu} = \pi$ and $\delta_{13} = \pi, \, \delta_{e\mu} = 0$, \eref{eq:deltaprob_emu_plusminus_num} gives $\Delta P_{e\mu}^+ \approx 0$, $\Delta P_{e\mu}^- \approx -0.007$. This is supported by numerical results, as shown in \fref{fig:overlapping_ellipses_nh_deltaminus}. (These are the ``up" and ``down" triangles. To represent values of $\delta_{e\mu}$ or $\delta_{e\tau}$, we use triangular plot markers whose orientation can be thought of as hands on a clock where 12 hours $\leftrightarrow 2\pi$.) Compared with the standard oscillation case, it is evident that the points with $\delta_+ = 0, \pi$ have moved $\sim \pm 0.01$ in the $P^-$ direction from the approximate position where $\delta_{13} = 0, \pi$ were located (\fref{fig:L1300_x0_SM_NH}).

For $\delta_{+} = \pi/2, 3\pi/2$ (the triangles pointing left and right in \fref{fig:overlapping_ellipses_nh_deltaminus}) the points are split in the $P_{e\mu}^+$ direction instead, as expected based on $\Delta P_{e\mu}^+ \sim -\sin\delta_+$, $\Delta P_{e\mu}^- \sim \cos\delta_+$. This is not specific to hidden sector ellipses with $\delta_{13} = 0, \pi$, as shown in \fref{fig:overlapping_ellipses_b_nh} for $\delta_{13} = \pi/4$, $3\pi/4$. We can see that the relationship suggested by the perturbative expressions \eref{eq:prob_zero_plusminus_num}, \eref{eq:deltaprob_emu_plusminus_num} and our above reasoning is approximate, but captures the behavior to within $\lesssim 0.01$ on the biprobability plot.

It is useful to consider properties of the ``hidden sector ellipses" of fixed $\delta_{13}$ and varying $\delta_{e\mu}$ or $\delta_{e\tau}$, such as their width or the location where they are centered. These may be estimated in a straightforward way based on \eref{eq:prob_zero_plusminus_num}, \eref{eq:deltaprob_emu_plusminus_num}, and \eref{eq:deltaprob_etau_plusminus_num}. The lowest-order terms have $\Delta P^+_{e\mu, \, e\tau} \propto \sin(\delta_+)$, $\Delta P^-_{e\mu, \, e\tau} \propto \cos(\delta_+)$, i.e. the form of an ellipse whose major (minor) axis is aligned with the $P^+$ ($P^-$) axis. As $\delta_{e\tau}$ varies with all other parameters fixed, points on the hidden sector ellipse vary between:
\begin{align}\label{eq:emu_width}
	\Delta P^+_{e\mu} & \approx \pm 0.21 |\epsilon_{e\mu}|, \nonumber \\
	\Delta P^-_{e\mu} & \approx \pm 0.14 |\epsilon_{e\mu}|
\end{align}
for nonzero $\epsilon_{e\mu}$ and
\begin{align}\label{eq:etau_width}
	\Delta P^+_{e\tau} & \approx \pm 0.21 |\epsilon_{e\tau}|, \nonumber \\
	\Delta P^-_{e\tau} & \approx \pm 0.13 |\epsilon_{e\tau}|
\end{align}
for nonzero $\epsilon_{e\tau}$, where the numerical values are obtained for $L = 1300$ km, $\rho = 3$ g/cm$^3$ and $\Delta_{31} = \pi/2$. The approximate center of the ellipse (with $\Delta P^{\pm}_{e\mu, \, e\tau} = 0$) is determined by \eref{eq:d13_degen}. Therefore, at the leading order the width of the hidden sector ellipses is determined by $\epsilon_{e\tau}$ and the center is determined by $\sin(\delta_{13})$. So one can say at this level of approximation that hidden sector ellipses of the same $\sin\delta_{13}$ are centered at the same point, and with the same $\epsilon_{e\mu}$ or $\epsilon_{e\tau}$ they have approximately the same major and minor axis, so they overlap.

It's interesting that the effect of NSI on probabilities depends so strongly on $\delta_+$, and we can gain some insight by noticing that the $\delta_{e\mu}$ or $\delta_{e\tau}$ terms in \eref{eq:deltaprob_emu_plusminus} or \eref{eq:deltaprob_etau_plusminus} are suppressed by a power of $\Delta m^2_{21} / \Delta m^2_{31}$. In the $\Delta m^2_{21} / \Delta m^2_{31} \rightarrow 0$ limit the number of independent phases that control oscillation probabilities is reduced (see e.g. the phase reduction theorem in Section IV.C of \cite{Kikuchi:2008vq}). In the absence of NSI, we can see from \eref{eq:prob_zero_plusminus} that taking $\Delta m^2_{21}/\Delta m^2_{31} \rightarrow 0$ eliminates the effects of $\delta_{13}$, so there would be no intrinsic CP violation. With only one nonzero NSI element $\epsilon_{e\mu}$ or $\epsilon_{e\tau}$, taking the limit $\Delta m^2_{21} / \Delta m^2_{31} \rightarrow 0$ would leave only one phase, $\delta_+ = \delta_{13} + \delta_{e\mu}$ or $\delta_{13} + \delta_{e\tau}$. Using the actual nonzero value of $\Delta m^2_{21}$, the individual phases $\delta_{e\mu}$ or $\delta_{e\tau}$ are not absent but suppressed by small but nonzero $\Delta m^2_{21} / \Delta m^2_{31}$. Therefore, the result that hidden sector ellipses of constant $\delta_{13}$, $\pi - \delta_{13}$ approximately overlap with a $\delta_+$ degeneracy can be thought of as a consequence of the mass hierarchy.\footnote{``Mass hierarchy" used in its original sense to mean difference in scales $|\Delta m^2_{21} / \Delta m^2_{31}| \ll 1$, not the common practice of referring to mass ordering, i.e. sign of $\Delta m^2_{31}$.}

\begin{figure}[t]
	\centering
		\includegraphics[width=0.45\textwidth]{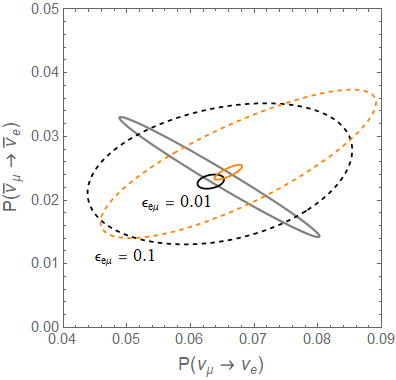}
\caption{\label{fig:vary_epsilon}Depending on the value of $\epsilon_{e\mu}$, the degeneracy may take different forms. For $\epsilon_{e\mu} \gtrsim 0.01$, there are approximately overlapping hidden sector ellipses. For $\epsilon_{e\mu} \sim 0.01$, the ``low $P^+$" side of one approximately overlaps the ``high $P^+$" side of the other. For $\epsilon_{e\mu} \ll 0.01$ the hidden sector ellipses are about as well-separated in the $P^+$ direction as the original $P_0^+$ values. In this case the value of $\delta_{e\mu}$ does not significantly interfere with determination of $\delta_{13}$.}
\end{figure}

The degeneracy depends on the magnitude $|\epsilon_{e\mu}|$ as well, as shown in \fref{fig:vary_epsilon}. If $\epsilon_{e\mu}$ is ``large enough" ($\gtrsim 0.01$), there are approximately overlapping hidden sector ellipses. (This is consistent with a similar observation in \cite{Agarwalla:2016fkh}.) For $\epsilon_{e\mu} \sim 0.01$, the ``low $P^+$" side of one approximately overlaps the ``high $P^+$" side of the other. For ``small enough" $\epsilon_{e\mu}$ ($\ll 0.01$) the hidden sector ellipses are about as well-separated in the $P^+$ direction as the original $P_0^+$ values. In this case the value of $\delta_{e\mu}$ does not significantly interfere with determination of $\delta_{13}$ (this is expected: as $\epsilon_{e\mu}\rightarrow 0$ in \eref{eq:deltaprob_emu_plusminus_num}, the NSI effects $\Delta P_{e\mu}^{\pm}$ go away).

It is evident from \eref{eq:prob_zero_plusminus_num}, \eref{eq:deltaprob_emu_plusminus_num} and \eref{eq:deltaprob_etau_plusminus_num} that the qualitative behavior we have just described does not change under $\Delta_{31} \rightarrow -\Delta_{31}$, which well approximates the difference between normal and inverted hierarchy. We have confirmed numerically that the same characteristics of degeneracies hold for the inverted hierarchy as suggested by the above reasoning. Furthermore, comparing the form of \eref{eq:deltaprob_emu_plusminus_num} with \eref{eq:deltaprob_etau_plusminus_num}, we can see that the behavior just described for $\epsilon_{e\mu} \neq 0$, $\epsilon_{e\tau} = 0$ should carry over to the case where $\epsilon_{e\mu} = 0$, $\epsilon_{e\tau} \neq 0$. This is also supported by numerical solutions; an example is shown in \fref{fig:overlapping_ellipses_c_nh}.

Thus, if one knows the presence of nonzero $\epsilon_{e\mu}$ or $\epsilon_{e\tau}$, there is at first order a degeneracy where only $\delta_+$ is determined. A further complication in the degeneracy picture is not knowing whether $\epsilon_{e\mu}$ or $\epsilon_{e\tau}$ is nonzero. We saw this in \fref{fig:ellipse_examples} as overlap between biprobability ellipses with and without NSI. More quantitatively, comparison of \fref{fig:L1300_x0_SM_NH} with \fref{fig:overlapping_ellipses_b_nh} (showing hidden sector ellipses for $\delta_{13} = \pi/4$ and $3\pi/4$) shows that the point on the biprobability plot corresponding to $\epsilon_{e\mu} = 0$ and $\delta_{13} = \pi/2$ may also correspond to $\epsilon_{e\mu} = 0.05$ and either $\delta_{13} = \pi/4$, $\delta_{e\mu} \sim 5\pi/4$ (black curve) or $\delta_{13} = 3\pi/4$, $\delta_{e\mu} \sim 7\pi/4$ (blue curve). Other points on these hidden sector curves intersect the $\epsilon_{e\mu} = 0$ curve near $\delta_{13} = 0$ or $\pi$ -- if nonzero $\epsilon_{e\mu}$ isn't suspected, this would simply look like there is little or no intrinsic CP violation in the leptonic sector. Returning to \eref{eq:deltaprob_emu_plusminus_num} and \eref{eq:deltaprob_etau_plusminus_num}, we can see that the $\sin\delta_{13}$ and $\cos\delta_+$ terms compete to determine the position of a hidden sector ellipse in the $P^-$ direction, while $\cos\delta_{13}$ and $\sin\delta_+$ compete in the $P^+$ direction. As we will see in \sref{sec:lifting_cp_degen}, the $\delta_+$ dependent terms help resolve this degeneracy as energy is varied.

\section{Lifting Phase Degeneracies}\label{sec:lifting_cp_degen}

As mentioned in \sref{sec:intro}, variation in oscillation probabilities with neutrino energy and baseline length can help break degeneracies, a situation we now study using the above results. For convenience and motivated by experiments like DUNE and T2HK, at a given baseline length $L$ we define $E_{0}$ as the neutrino energy for which $\Delta_{31} = \pi/2$. Then an arbitrary neutrino energy can be parametrized as
\begin{align}\label{eq:x_definition}
	E & = E_{0}\left( 1 + x \right)
\end{align}
so the parameter $x \equiv (E - E_0)/E_0$ is the fractional difference in energy from $\Delta_{31} = \pi/2$. To get a sense of the numbers involved, for a situation where $\Delta_{31} = \pi/2$ at an energy of 2.5 GeV, the parameter $x$ will vary between $x = \pm 0.2$ as energy varies from 2.0 GeV to 3.0 GeV. This is motivated by DUNE, which will have a wide band of energies. By comparison, a narrow band beam like NO$\nu$A is often well approximated as a single energy, as in \cite{Friedland:2012tq} for example. With this in mind, we will first expand to lowest order in $x$ about $x=0$ (i.e. expand about $E = E_0$) terms in $P$ and $\overline{P}$ that depend on $\Delta_{31}$. Then we will compare this with the behavior near $x = -\frac{2}{3}$, motivated by the off-axis beams proposed for T2HKK.

\subsection{Behavior Around $\Delta_{31} \sim \pi/2$ \, ($x \sim 0$)}\label{sec:x0_behavior}

Based on the definition \eref{eq:x_definition}, we can make the exact substitutions
\begin{align}\label{eq:delta31_parametrized}
	\Delta_{31} & = \frac{\Delta m^2_{31} L}{4 \hbar c E_{0}} \, \frac{1}{1 + x} \, = \, \frac{\pi}{2} \, \frac{1}{1 + x},\\
	{\rm or} \, \, \, \, \frac{1}{\Delta_{31}} & = \frac{2}{\pi}(1 + x),
\end{align}
and expand to first order in $x$. Then \eref{eq:prob_zero_plusminus_num} becomes
\begin{align}\label{eq:prob_zero_plusminus_x}
	P_0^+ & \approx 0.059 + 0.0013 \, \left( 1 - 2x \right) + 0.027 x\cos(\delta_{13}) \nonumber \\
	P_0^- & \approx -0.018 \, \left( 1 - x \right)\sin(\delta_{13}) + \left( 1.1\times 10^{-5} \right) \, L \, \rho \, \left( 0.64 - 0.93 x \right),
\end{align}
\eref{eq:deltaprob_emu_plusminus_num} becomes
\begin{align}\label{eq:deltaprob_emu_plusminus_x}
	\Delta P_{e\mu}^+ & = \left( 5.4 \times 10^{-5} \right) \, L \, \rho \, |\epsilon_{e\mu}| \left[ -\sin(\delta_+) + 0.099 \sin(\delta_{e\mu}) \right], \nonumber \\
	\Delta P_{e\mu}^- & = \left( 5.6\times 10^{-5} \right) \, L \, \rho \, |\epsilon_{e\mu}| \left[ \left( 0.64 - 0.84 x \right)\cos(\delta_+) + \left(0.14 + 0.0077 x \right) \cos(\delta_{e\mu})\right],
\end{align}
and \eref{eq:deltaprob_etau_plusminus_num} becomes
\begin{align}\label{eq:deltaprob_etau_plusminus_x}
	\Delta P_{e\tau}^+ & = \left( 5.4\times 10^{-5} \right) \, L \, \rho \, |\epsilon_{e\tau}|\left[ \sin(\delta_+) + 0.096 \sin(\delta_{e\tau}) \right], \nonumber \\
	\Delta P_{e\tau}^- & = \left( 5.4 \times 10^{-5} \right) \, L \, \rho \, |\epsilon_{e\tau}| \left[ \left( 0.64 - 0.93 x \right)\cos(\delta_+) - \left( 0.15 - 0.30 x \right) \cos(\delta_{e\tau}) \right].
\end{align}

We can see from \eref{eq:prob_zero_plusminus_x} that the $\delta_{13}$, $\pi - \delta_{13}$ degeneracy is broken by increasing the separation in the $P^+$ direction, since $\sin(\delta_{13}) = \sin(\pi-\delta_{13})$ but $\cos(\delta_{13}) = -\cos(\pi-\delta_{13})$, so the cosine term in $P^+$ is responsible for separating these points based on $\delta_{13}$. For instance, $\delta_{13} = 0, \pi$ gives $\cos\delta_{13} = \pm 1$, so that $x \neq 0$ moves these in opposite directions along $P^+$. Furthermore, since representative numbers give $\left( 5.4 \times 10^{-5} \right) \, (1300) \, (3) \, (0.05) \approx 0.011$ (and correspondingly less for smaller $\epsilon$ or $L < 1300$ km), it is the standard $\delta_{13}$ terms that dominate as the energy changes. Comparison with \fref{fig:L1300_xm0p2_SM_NH} shows the advantage and limitations of the perturbative solutions \eref{eq:prob_zero_plusminus_x}, \eref{eq:deltaprob_emu_plusminus_x}, \eref{eq:deltaprob_etau_plusminus_x}. The qualitative behavior agrees, with the $\delta_{13}$ degeneracy is broken by stretching the ellipse in the $P^+$ direction, but as expected with a first-order perturbative expansion, there are clearly higher-order effects missing that prevent the precise description of this behavior.

\begin{figure}[t]
	\centering
	\begin{subfigure}[b]{0.23\textwidth}
		\centering
		\includegraphics[width=\textwidth]{overlapping_ellipses_nh_a.png}
		\caption{\label{fig:overlapping_ellipses_nh_deltaminus_2}}
	\end{subfigure} \, \, %
	\begin{subfigure}[b]{0.23\textwidth}
		\centering
		\includegraphics[width=\textwidth]{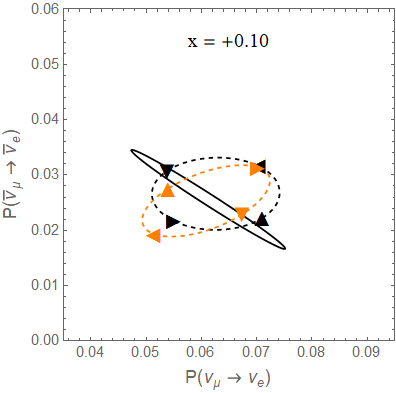}
		\caption{\label{fig:overlapping_ellipses_nh_deltaminus_x0p1}}
	\end{subfigure} \, \, %
	\begin{subfigure}[b]{0.23\textwidth}
		\centering
		\includegraphics[width=\textwidth]{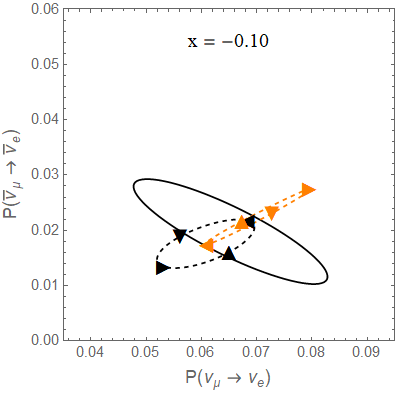}
		\caption{\label{fig:overlapping_ellipses_nh_deltaminus_xm0p1}}
	\end{subfigure} \, \, %
	\begin{subfigure}[b]{0.23\textwidth}
		\centering
		\includegraphics[width=\textwidth]{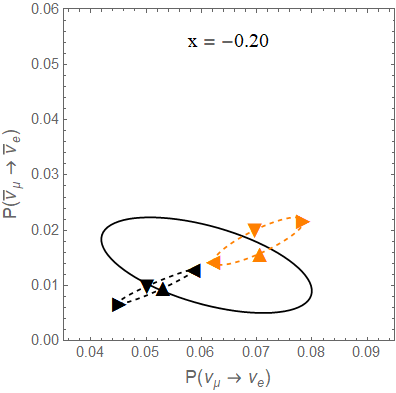}
		\caption{\label{fig:overlapping_ellipses_nh_deltaminus_xm0p2}}
	\end{subfigure}
\caption{\label{fig:ellipse_variation_offpeak_nh}Numerical solutions illustrating the same features as the perturbation result \eref{eq:prob_zero_plusminus_x} and \eref{eq:deltaprob_emu_plusminus_x}. \fref{fig:overlapping_ellipses_nh_deltaminus_2} shows the ``hidden sector" biprobability curves for the normal hierarchy with $L=1300$ km and $\Delta_{31} = \pi/2$. \fref{fig:overlapping_ellipses_nh_deltaminus_x0p1} shows the situation with energy increased from the peak by a fractional amount $x = 0.1$, while \fref{fig:overlapping_ellipses_nh_deltaminus_xm0p1} and \fref{fig:overlapping_ellipses_nh_deltaminus_xm0p2} show the energy reduced from the peak, they correspond to fractional changes of $x = -0.1$ and $x = -0.2$ respectively.}
\end{figure}

\begin{figure}[h]
	\centering
	\begin{subfigure}[b]{0.23\textwidth}
		\centering
		\includegraphics[width=\textwidth]{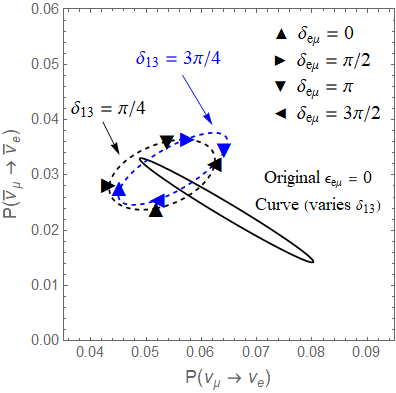}
		\caption{\label{fig:overlapping_ellipses_b_nh}}
	\end{subfigure} \, \, %
	\begin{subfigure}[b]{0.23\textwidth}
		\centering
		\includegraphics[width=\textwidth]{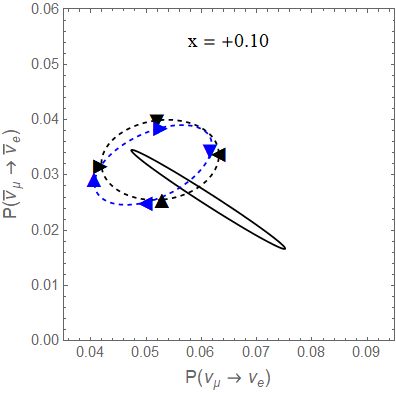}
		\caption{\label{fig:overlapping_ellipses_b_nh_x0p1}}
	\end{subfigure} \, \, %
	\begin{subfigure}[b]{0.23\textwidth}
		\centering
		\includegraphics[width=\textwidth]{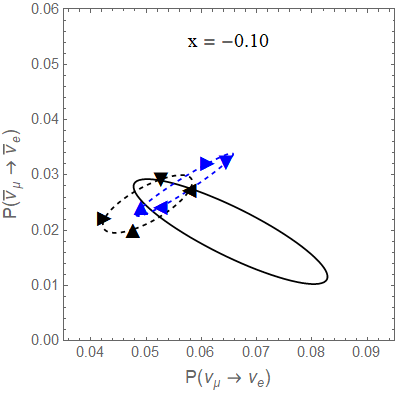}
		\caption{\label{fig:overlapping_ellipses_b_nh_xm0p1}}
	\end{subfigure} \, \, %
	\begin{subfigure}[b]{0.23\textwidth}
		\centering
		\includegraphics[width=\textwidth]{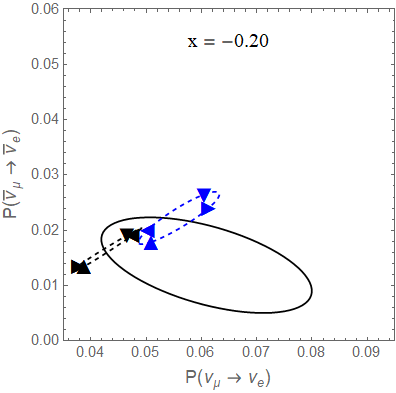}
		\caption{\label{fig:overlapping_ellipses_b_nh_xm0p2}}
	\end{subfigure}
\caption{\label{fig:ellipse_variation_offpeak_nh_b}Biprobability plots for $\delta_{13} = \pi/4$ and $\delta_{13} = 3\pi/4$ at $\Delta_{31} = \pi/2$ (\fref{fig:overlapping_ellipses_b_nh}) are approximately degenerate, but increasing (\fref{fig:overlapping_ellipses_b_nh_x0p1}) or decreasing (\fref{fig:overlapping_ellipses_b_nh_xm0p1} and \fref{fig:overlapping_ellipses_b_nh_xm0p2}) the energy by fractional amounts $x = \pm 0.1$ or $-0.2$ works to lift the degeneracy.}
\end{figure}

It is also evident from the perturbative results that decreasing the energy, so that the fraction $x$ is negative, enhances the effect of $\sin\delta_{13}$, $\cos\delta_+$, and the matter effect due to the $\sim (1-x)$ terms, while increasing the energy by a fraction will diminish these terms. This means that the energy spectrum below $E_0$ should be most useful in breaking degeneracies; this is also seen numerically in \fref{fig:ellipse_variation_offpeak_nh} and \fref{fig:ellipse_variation_offpeak_nh_b}. The influence of the $x\cos\delta_{13}$ term in $P_0^+$ is evident, as $x<0$ moves the $\delta_{13} = 0$ curve toward lower $P^+$ and the $\delta_{13} = \pi$ curve to higher $P^+$ relative to each other. \fref{fig:ellipse_variation_offpeak_nh} shows for the normal hierarchy how NSI ellipses with constant $\delta_{13} = 0$ and $\delta_{13} = \pi$ at $\Delta_{31} = \pi/2$ (\fref{fig:overlapping_ellipses_nh_deltaminus_2}) are affected as the energy varies: increased by $x = +0.1$ in \fref{fig:overlapping_ellipses_nh_deltaminus_2}, and decreased by $x = -0.1$ in \fref{fig:overlapping_ellipses_nh_deltaminus_x0p1} and $x = -0.2$ in \fref{fig:overlapping_ellipses_nh_deltaminus_xm0p2}). \fref{fig:ellipse_variation_offpeak_nh_b} shows the same result for $\delta_{13} = \pi/4$ and $3\pi/4$.

\begin{figure}[h]
	\centering
	\begin{subfigure}[b]{0.33\textwidth}
		\centering
		\includegraphics[width=\textwidth]{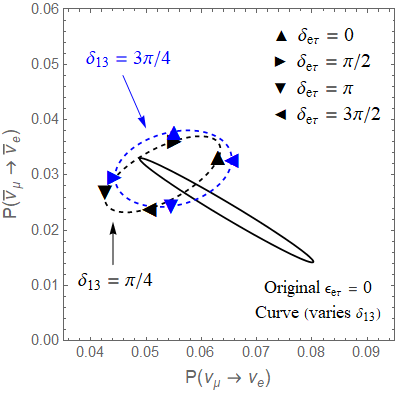}
		\caption{\label{fig:overlapping_ellipses_c_nh}}
	\end{subfigure} \, \, \, \, \, \, \, \, \, \, %
	\begin{subfigure}[b]{0.33\textwidth}
		\centering
		\includegraphics[width=\textwidth]{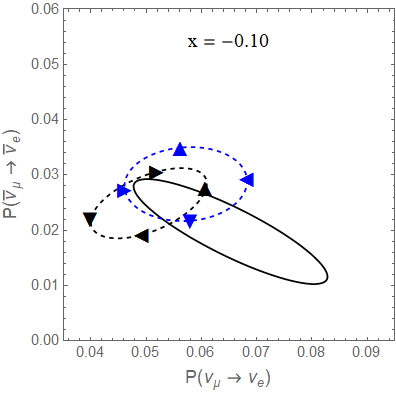}
		\caption{\label{fig:overlapping_ellipses_c_nh_xm0p1}}
	\end{subfigure}
\caption{\label{fig:ellipse_variation_offpeak_nh_c}Degeneracies and breaking are qualitatively the same for $\epsilon_{e\tau}$ as for $\epsilon_{e\mu}$, as suggested by the similar form of their perturbative expressions. This figure shows biprobability hidden sector ellipses for $\epsilon_{e\tau} = 0.05$, $\delta_{13} = \pi/4$ and $\delta_{13} = 3\pi/4$ at $\Delta_{31} = \pi/2$ (\fref{fig:overlapping_ellipses_c_nh}) are approximately degenerate, but changing the energy by fractional amounts $x = \pm 0.1$ or $-0.2$ works to lift the degeneracy (\fref{fig:overlapping_ellipses_c_nh_xm0p1}).}
\end{figure}

\fref{fig:ellipse_variation_offpeak_nh_c} shows a similar result for hidden sector ellipse degeneracy with $\epsilon_{e\tau} = 0.05$. \fref{fig:overlapping_ellipses_c_nh} has $\Delta_{31} = \pi/2$ and the hidden sector ellipses that vary $\delta_{e\tau}$ for fixed $\delta_{13} = \pi/4, 3\pi/4$ are approximately degenerate. As in the $\epsilon_{e\mu}$ case, and as suggested by \eref{eq:deltaprob_etau_plusminus_x}, variation in the energy works to lift the degeneracy. This is shown in \fref{fig:overlapping_ellipses_c_nh_xm0p1}, where the energy has been shifted by $x = -0.1$.

Finally, we point out how variations in $P^{\pm}$ relate back to the spectrum of probabilities (or number of events at the detector) as a function of neutrino energy. Neutrinos and antineutrinos having identical spectra would mean that for any energy, the location on the biprobability plot is along $P^+$ (i.e. on the $P^- = 0$ axis). Increasing along the $P^+$ direction means that the height of both neutrino and antineutrino spectra are increased equally, while increasing (decreasing) along the $P^-$ direction means that for the given energy, the probabilities in the neutrino spectrum are increased (decreased) relative to the antineutrino spectrum.

\subsection{Behavior Around $\Delta_{31} \sim 3\pi/2$ \, ($x \sim -2/3$)}\label{sec:x23_behavior}

Even within a broad neutrino energy spectrum around $\Delta_{31} = \pi/2$ with $x$ between $\sim \pm 0.2$, it is evidently difficult to separate points such as $\epsilon_{e\tau} = 0$, $\delta_{13} = -\pi/2$ and $\epsilon_{e\tau} = 0.2$, $\delta_{13} = \delta_{e\tau} = 0$. We can improve on this by thinking about the various terms in \eref{eq:prob_zero_plusminus_num} and \eref{eq:deltaprob_etau_plusminus_num} change when we look at $\Delta_{31} = 3\pi/2$ instead of $\pi/2$. The terms proportional to $\Delta_{31} = \frac{\pi}{2}\frac{1}{1+x}$ are accentuated (increased from $\pi/2$ to $3\pi/2$) and the terms proportional to $1/\Delta_{31}$ are proportionally decreased. At $x = 0$ the term $\sin(\Delta_{31}) = 1$ and at $x = -2/3$ the quantity $\sin(\Delta_{31}) = \sin(\frac{\pi}{2} \frac{1}{1+x}) = -1$, so the $\sin(\Delta_{31})^2$ terms are unaffected. Collectively, this has the effect on $P^-$ of making $\delta_{e\tau}$ from subdominant to dominant compared with $\delta_+$. The $P^+$ term is unaffected.

The above reasoning implies that for $x \sim -2/3$, i.e. $\Delta_{31} \sim 3\pi/2$, points would be well-separated in the $P^-$ direction based on $\delta_{e\tau}$, with only a subdominant effect from $\delta_+$. In \sref{sec:example} we will put this to use for the study of the degeneracy at apparent $\delta_{13} \sim -\pi/2$. \fref{fig:maxcpv_degen_unjang} shows numerical results that support the analytic development here.

A relevant situation is a long-baseline experiment with detectors off-axis by a few degrees, such as the proposed T2HKK sites \cite{Abe:2016ero}. We can evaluate these proposed T2HKK sites using the approximate relation $E_{\nu}^{\rm peak} \approx \left( 30 \, {\rm MeV}\right) / \theta$ for the peak neutrino energy at off-axis angle $\theta$ \cite{Beavis:1995pbs,Itow:2001ee,Ayres:2002ws,McDonald:2001mc}. The Unjang and Minjuji sites have $\Delta_{31} = 1.51\pi$, $1.59\pi$ respectively ($x = -0.679$ and $-0.694$); in contrast, the Bisul site has $\Delta_{31} = 0.82\pi$ (or $x = -0.405$). This means that the Unjang and Minjuji sites can separate points further in biprobability space according to $\delta_{e\tau}$, in contrast with separation based on $\delta_+$ as we have seen so far. (\fref{fig:maxcpv_degen_unjang} referenced above shows results for parameters relevant to the Unjang site.) It is worth noting that the overall neutrino flux decreases with increased off-axis angle. This is an important consideration in evaluating a specific experiment's ability to distinguish two parameter choices, but we leave it for future work as it is beyond the scope outlined in \sref{sec:intro}.

\section{Degeneracies for Apparent $\delta_{13} \sim -\pi/2$}\label{sec:example}

The results of the previous sections may now be applied to gain insight into a case of current interest: the implication of probabilities that suggest $\delta_{13} \approx -\pi/2$. While no values of $\delta_{13}$ have been ruled out experimentally, recent results from the T2K and NO$\nu$A experiments favor $\delta_{13} \neq 0, \pi$. T2K gives $\delta_{13}$ in the range $1.06\pi$ to $1.86\pi$ at $90\%$ confidence level \cite{Abe:2017vif}, while NO$\nu$A gives $\delta_{13}$ in the range 0 to $0.12\pi$ or $0.91\pi$ to $2\pi$ at $68.3\%$ confidence level, with best fit $1.21\pi$ \cite{NOvA:2018gge}. As previously stated, this paper does not aim for precise analysis of any experiment, but we will take this as motivation to specifically point out the implications of our work for the $P,\overline{P}$ points centered around $\delta_{13} = -\pi/2$. Earlier results consistent with these have been discussed in \cite{Forero:2016cmb}, which shows a few examples of the biprobability curves that we have called ``hidden sector" curves, corresponding to $\epsilon_{e\tau} = 0.3$ and $\delta_{13} = 0$ or $\pi$. They concluded that this particular choice of parameters is consistent with T2K and NO$\nu$A results, i.e. there is a degeneracy of the type we have considered in this paper. This section presents results consistent with \cite{Forero:2016cmb} for varying $\epsilon_{e\tau}$ and $\delta_{13}$, and also considers degeneracy breaking. Other work has considered the implications of apparent $\delta_{13} = -\pi/2$ for NSI and the possibility of sterile neutrino mixing \cite{Palazzo:2015gja}.

Recalling our previous discussion of the approximate $\delta_{13},$ $\pi - \delta_{13}$ degeneracy in the standard oscillation case, the ``endpoints" of the narrow biprobability ellipse at $\delta_{13} = \pm\pi/2$ are special cases because the approximate $\delta_{13}$, $\delta_{13} - \pi$ degeneracy goes away. However, the possibility of $\epsilon_{e\tau} \neq 0$ means the other degeneracies we've considered may still apply. Specifically, a hidden sector ellipse for some $\delta_{13} \neq -\pi/2$ may overlap with point where the standard oscillation case predicts $\delta_{13} = -\pi/2$. And as we've seen, this means that the $\pi - \delta_{13}$ hidden sector ellipse would also approximately overlap at this point. The larger $\epsilon_{e\tau}$ is, the farther this $\delta_{13}$ or $\pi - \delta_{13}$ would be from the apparent value of $\delta_{13} = -\pi/2$.
\begin{figure}[t]
	\centering
	\begin{subfigure}[b]{0.35\textwidth}
		\centering
		\includegraphics[width=\textwidth]{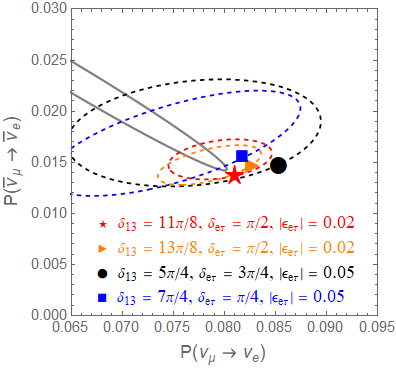}
		\caption{\label{fig:degen_3pi2}}
	\end{subfigure}\\
	\begin{subfigure}[b]{0.35\textwidth}
		\centering
		\includegraphics[width=\textwidth]{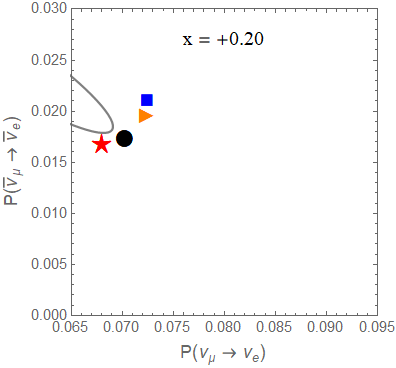}
		\caption{\label{fig:degen_3pi2_x02}}
	\end{subfigure} \, \, \, \, \, \, \, \, \, \, %
	\begin{subfigure}[b]{0.35\textwidth}
		\centering
		\includegraphics[width=\textwidth]{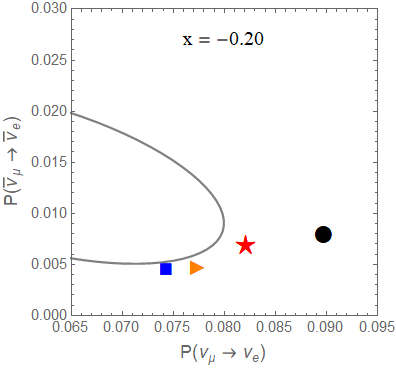}
		\caption{\label{fig:degen_3pi2_xm02}}
	\end{subfigure}
\caption{\label{fig:3pi2_degen_breaking}Examples of degeneracy at the point where $\delta_{13} = 3\pi/2$ in the standard 3-neutrino picture.}
\end{figure}

This is illustrated in \fref{fig:degen_3pi2} for $\epsilon_{e\tau} = 0.02, \, 0.05$. On each hidden sector curve, we've marked one specific point that happens to be close to the $\delta_{13} = -\pi/2$ point of the standard standard oscillation case. In general, it's evident that larger $\epsilon_{e\tau}$ directly corresponds to larger deviation of $\delta_{13}$ from the apparent value. In this case, all marked points have $\delta_+ \sim 2\pi \rightarrow 0$, so that none of the $\delta_+$ dependent terms in the perturbation expansion help us distinguish the points by varying energy. Furthermore, the $\delta_{13}$'s we consider are all within $\pi/4$ of the original point in question ($\delta_{13} = -\pi/2$). (Larger $\epsilon_{e\tau}$ improves this situation.) This means the $\delta_{13}$ dependent terms, which are the dominant terms in the perturbation expansion, are also restricted in their use.

Varying the energy still does help break the degeneracy, as seen in \fref{fig:degen_3pi2_x02} where the energy is increased by $x = +0.2$, and in \fref{fig:degen_3pi2_xm02} where the energy is decreased by $x = -0.2$. Looking at energies below the $E_0$ where $\Delta_{31} = \pi/2$ is still helpful in separating the points, but the close proximity of all of these points in parameter space of CP phases means that the first-order approximation we've used is less helpful. We leave the consideration of higher-order terms and their effect on this picture to future work.

\begin{figure}[t]
	\centering
	\begin{subfigure}[b]{0.35\textwidth}
		\centering
		\includegraphics[width=\textwidth]{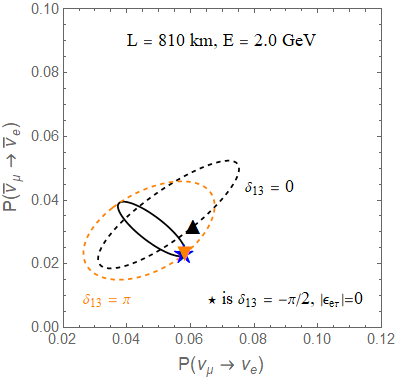}
		\caption{\label{fig:maxcpv_degen_nova}}
	\end{subfigure} \, \, \, \, \, \, \, \, \, \, %
	\begin{subfigure}[b]{0.35\textwidth}
		\centering
		\includegraphics[width=\textwidth]{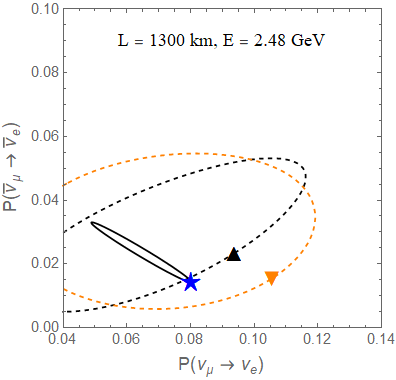}
		\caption{\label{fig:maxcpv_degen_dune}}
	\end{subfigure}\\
	\begin{subfigure}[b]{0.35\textwidth}
		\centering
		\includegraphics[width=\textwidth]{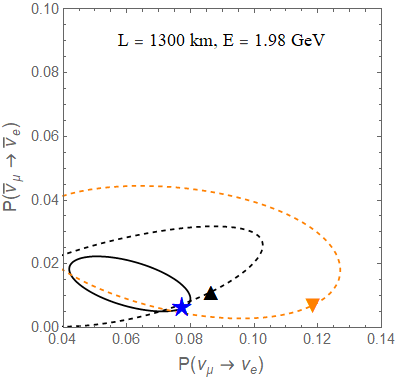}
		\caption{\label{fig:maxcpv_degen_dune_xm0p2}}
	\end{subfigure} \, \, \, \, \, \, \, \, \, \, %
	\begin{subfigure}[b]{0.35\textwidth}
		\centering
		\includegraphics[width=\textwidth]{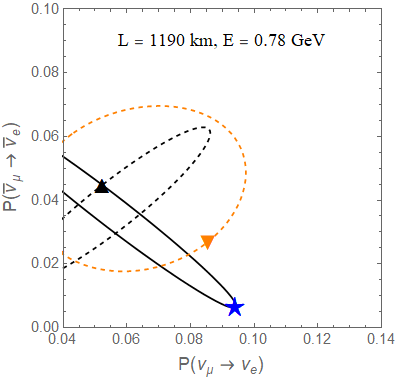}
		\caption{\label{fig:maxcpv_degen_unjang}}
	\end{subfigure}
\caption{\label{fig:maxcpv_nova_dune}Degeneracy for apparent $\delta_{13} = -\pi/2$ at the baseline and energy of NO$\nu$A. Here the star indicates $\delta_{13} = -\pi/2$ with no NSI, the triangle pointing up indicates $\delta_{13} = 0$, $|\epsilon_{e\tau}| = 0.2$, $\delta_{e\tau} = 0$, and the triangle pointing down indicates $\delta_{13} = \pi$, $|\epsilon_{e\tau}| = 0.2$, $\delta_{e\tau} = \pi$.  This is somewhat broken by the baseline and energy of DUNE (for $\Delta_{31} = \pi/2$, i.e. $x=0$) and further by looking towards $x = -0.2$. \fref{fig:maxcpv_degen_unjang} shows that the baseline length and energy of the proposed off-axis T2HKK site at Unjang further improves the split.}
\end{figure}

An interesting situation is shown in \fref{fig:maxcpv_nova_dune}. A measurement that apparently gives maximal CP violation in the standard picture would be consistent with nonzero NSI and absence of CP violation in {\it both} the standard and hidden sectors. For apparent $\delta_{13} = -\pi/2$ at the baseline and energy of NO$\nu$A, the points for $|\epsilon_{e\tau}| = 0.2$, $\delta_{13} = \pi$, $\delta_{e\tau} = \pi$ and $|\epsilon_{e\tau}| = 0.2$, $\delta_{13} = 0$, $\delta_{e\tau} = 0$ are essentially degenerate with the no-NSI $\delta_{13} = -\pi/2$ result. At the baseline and energy of DUNE (with $x=0$), these points separate by an amount $\sim 0.01$ in the $P$ direction. However, looking at $x = -0.2$, we see that $\delta_{13}, \delta_{e\tau} = \pi$ is well separated from the other two points (which are not distinguished according to the $\Delta P \lesssim 0.01$ rule of thumb referenced in \sref{sec:degens}).

This can be understood using \eref{eq:prob_zero_plusminus_num} and \eref{eq:deltaprob_etau_plusminus_num}. At $\Delta_{31} = \pi/2$, the expressions for $P^{\pm} = P_0^{\pm} + \Delta P^{\pm}_{e\tau}$ approximately coincide for the above three parameter choices. Looking at the coefficients of $x$ in the expressions will then reveal how varying the energy does or does not break this degeneracy. In this case, it's convenient to return to the un-rotated coordinate $P = \frac{1}{\sqrt{2}}\left( P^+ + P^- \right)$ (using the definition \eref{eq:pplus_pminus_def}). Schematically, this expression looks like $P = \left( \cdots \right) + \frac{1}{\sqrt{2}} x \left( \cdots \right)$. For $\delta_{13} = -\pi/2$ and $\epsilon_{e\tau} = 0$ the term in parentheses multiplying $x$ gives $-0.058$, while for $\delta_{13} = 0$, $|\epsilon_{e\tau}| = 0.2$, $\delta_{e\tau} = 0$ this term has the similar value $-0.053$. However, for $\delta_{13} = \pi$, $|\epsilon_{e\tau}| = 0.2$, $\delta_{e\tau} = \pi$ this term gives $-0.11$, approximately twice the value for the other two. This is consistent with \fref{fig:maxcpv_nova_dune}, where varying energy so that $x < 0$ splits the point corresponding to $\delta_{13}, \delta_{e\tau} = \pi$ away from the other two along the $P$ direction. Taken together, the numerical results and the perturbation expansion illustrate the role of a broad energy spectrum in addition to different baseline lengths in breaking specific degeneracies.

Finally, we can draw on the results of \sref{sec:x23_behavior} to see how the degeneracy between $\delta_{13} = 0$, $\delta_{e\tau} = 0$ and $\delta_{13} = \pi/2$, $|\epsilon_{e\tau}| = 0$ may be further broken. Off-axis neutrino beams at an experiment like T2HKK can provide $\Delta_{31} = 3\pi/2$ (or $x = -2/3$), which provides the maximal separation along the $P^-$ direction of biprobability space. This is clearly seen in \fref{fig:maxcpv_degen_unjang}, which uses $L = 1190$ km, $E = 0.78$ GeV (relevant to proposed T2HKK Unjang site): the case $\delta_{13} = -\pi/2$, $|\epsilon_{e\tau}| = 0$ is now far-separated along the $P^-$ direction from $\delta_{13} = 0$, $\delta_{e\tau} = 0$. To reiterate the conclusion of \sref{sec:x23_behavior}: when switching from $\Delta_{31} \sim \pi/2 \, \rightarrow \, 3\pi/2$, the terms in $P^-$ switch from being dominated by $\delta_+$ to $\delta_{e\tau}$. We also reiterate that off-axis beams have a lower neutrino flux, so this further separation in biprobability space does not automatically mean easier resolution. In this paper we have focused on degeneracies in terms of the question ``when do different parameters give similar predictions?" and note that this is a case where very different parameters (no CP violation vs. maximal CP violation) give the same prediction at some experiments, and quite different predictions at another.


\section{Summary and Future Work}\label{sec:conclusion}

Here we summarize the main features found in the body of this paper, and discuss future directions that aim to use this framework. In the absence of NSI, a given $\delta_{13}$ leads to probabilities approximately degenerate with $\pi - \delta_{13}$ when the baseline and neutrino energy are chosen to give $\Delta_{31} = \pi/2$. For fixed baseline, looking at neutrino energies away from the $E_0$ giving $\Delta_{31} = \pi/2$ works to break this degeneracy. On a biprobability plot, decreasing the energy ($x<0$) from $E_0$ works best to increase separation in the $P^+$ direction, improving the possibility for resolving different parameters.

In the presence of nonzero NSI parameter $\epsilon_{e\mu}$ (or $\epsilon_{e\tau}$), the effect on the oscillation probabilities near $\Delta_{31} = \pi/2$ is dominated by the sum $\delta_+ = \delta_{13} + \delta_{e\mu}$ (or $\delta_{13} + \delta_{e\tau}$), with a smaller contribution from $\delta_{e\mu}$ (or $\delta_{e\tau}$) alone. In combination with the first result listed above there is an approximate degeneracy among pairs of ``hidden sector" ellipses with $\delta_{13}$ and $\pi - \delta_{13}$, as seen in \fref{fig:ellipse_variation_offpeak_nh} and \fref{fig:ellipse_variation_offpeak_nh_b} for $\epsilon_{e\mu}$, and \fref{fig:ellipse_variation_offpeak_nh_c} for $\epsilon_{e\tau}$. At leading order, these hidden sector ellipses have a center determined by $\sin(\delta_{13})$ and major/minor axis length determined by $|\epsilon_{e\mu}|$ or $|\epsilon_{e\tau}|$. As in the no-NSI case, looking at neutrino energies below the $E_0$ corresponding to $\Delta_{31} = \pi/2$ tends to increase separation between two previously-degenerate points in the $P^+$ direction of a biprobability plot. These results are consistent with previous work on degeneracies in the presence of nonzero $\epsilon_{e\tau}$ \cite{Liao:2016hsa}, but provide further insight into how individual parameters affect the degeneracy as presented in biprobability space.

For probabilities consistent with $\delta_{13} \sim -\pi/2$ in the standard oscillation case, nonzero $\epsilon_{e\tau}$ can produce the same oscillation probabilities for certain values of $\delta_{13}$ and $\delta_{e\tau}$. The larger $\epsilon_{e\tau}$ is, the more $\delta_{13}$ can vary away from the apparent value of $-\pi/2$. In particular, apparent $\delta_{13} = -\pi/2$ (maximal CP violation in the standard case) is also consistent with $|\epsilon_{e\tau}| = 0.2$ and $\delta_{13}, \delta_{e\tau} = 0$ or $\delta_{13}, \delta_{e\tau} = \pi$ (NSI with no CP violation). For experimental parameters of DUNE, these points are more easily distinguishable than at NO$\nu$A. In particular, both numerical results and the perturbative approach in this paper show that the $\delta_{13}, \delta_{e\tau} = \pi$ point becomes well-separated as energy varies, but the $\delta_{13}, \delta_{e\tau} = 0$ point remains more difficult to distinguish from maximal CP violation in the standard scenario. This is consistent with results presented in \cite{Friedland:2012tq,Forero:2016cmb}, but specifically addresses the degeneracy between maximal, standard CP violation and NSI without CP violation while providing a clear relationship between the underlying parameters and their effect on degeneracies as energy is varied.

Furthermore, we have shown that baseline and energy combinations relevant for long-baseline off-axis beams like T2HKK may dramatically help split degenerate points along the $P^-$ direction, including points not distinguishable just from the energy spectrum at an experiment like DUNE. In the earlier terminology where $x$ is the fractional change in energy from the $\Delta_{31} = \pi/2$ maximum, such an off-axis experiment may provide $x \approx -2/3$. In comparison with $x \approx 0$, the relative importance of the $\delta_+$ and the $\delta_{e\tau}$ terms in the expression for $P^-$ is reversed.

The perturbative solutions have provided insight that supports each of these observations, but the first-order solutions presented in \cite{Arafune:1997hd,Cervera:2000kp,Kikuchi:2008vq} only take us so far. It may also be useful examine the next-order terms in \cite{Kikuchi:2008vq}, but this is probably not consistent as long the assumption that $\theta_{13}$ is small remains. Removing this assumption leads to perturbative expressions that can adjust the previous results by $\sim 0.001$ to $0.01$ for the $L$ and $E$ that interest us, and this is the same order as some effects we've considered here \cite{Asano:2011nj}.

This paper has considered approximate degeneracies and discussed how looking at varying energies tends to break degeneracies as viewed on a biprobability plot. This is a valid approximation for now because the difference between the black and blue curves in \fref{fig:overlapping_ellipses_c_nh}, for example, is much finer than any current experiment can distinguish. To help give context to the implications of how far separated biprobablility points are, we have roughly thought about a $\lesssim 0.01$ separation in biprobability space as being degenerate. However, we have specifically avoided precise statements that characterize exactly when two parameters are degenerate because the answer depends on further experimental details. A thorough examination of this for experiments such as DUNE and T2HKK is an important next step to build on this work.

\acknowledgments

I am very grateful to Cecilia Lunardini, Tanmay Vachaspati, Pilar Coloma, and Henry Lamm for useful conversations during early stages of this work.

\bibliography{refs}

\begin{thebibliography}{48}%
\makeatletter
\providecommand \@ifxundefined [1]{%
 \@ifx{#1\undefined}
}%
\providecommand \@ifnum [1]{%
 \ifnum #1\expandafter \@firstoftwo
 \else \expandafter \@secondoftwo
 \fi
}%
\providecommand \@ifx [1]{%
 \ifx #1\expandafter \@firstoftwo
 \else \expandafter \@secondoftwo
 \fi
}%
\providecommand \natexlab [1]{#1}%
\providecommand \enquote  [1]{``#1''}%
\providecommand \bibnamefont  [1]{#1}%
\providecommand \bibfnamefont [1]{#1}%
\providecommand \citenamefont [1]{#1}%
\providecommand \href@noop [0]{\@secondoftwo}%
\providecommand \href [0]{\begingroup \@sanitize@url \@href}%
\providecommand \@href[1]{\@@startlink{#1}\@@href}%
\providecommand \@@href[1]{\endgroup#1\@@endlink}%
\providecommand \@sanitize@url [0]{\catcode `\\12\catcode `\$12\catcode
  `\&12\catcode `\#12\catcode `\^12\catcode `\_12\catcode `\%12\relax}%
\providecommand \@@startlink[1]{}%
\providecommand \@@endlink[0]{}%
\providecommand \url  [0]{\begingroup\@sanitize@url \@url }%
\providecommand \@url [1]{\endgroup\@href {#1}{\urlprefix }}%
\providecommand \urlprefix  [0]{URL }%
\providecommand \Eprint [0]{\href }%
\providecommand \doibase [0]{http://dx.doi.org/}%
\providecommand \selectlanguage [0]{\@gobble}%
\providecommand \bibinfo  [0]{\@secondoftwo}%
\providecommand \bibfield  [0]{\@secondoftwo}%
\providecommand \translation [1]{[#1]}%
\providecommand \BibitemOpen [0]{}%
\providecommand \bibitemStop [0]{}%
\providecommand \bibitemNoStop [0]{.\EOS\space}%
\providecommand \EOS [0]{\spacefactor3000\relax}%
\providecommand \BibitemShut  [1]{\csname bibitem#1\endcsname}%
\let\auto@bib@innerbib\@empty
\bibitem [{\citenamefont {Branco}\ \emph {et~al.}(2012)\citenamefont {Branco},
  \citenamefont {Felipe},\ and\ \citenamefont {Joaquim}}]{Branco:2011zb}%
  \BibitemOpen
  \bibfield  {author} {\bibinfo {author} {\bibfnamefont {G.~C.}\ \bibnamefont
  {Branco}}, \bibinfo {author} {\bibfnamefont {R.~G.}\ \bibnamefont {Felipe}},
  \ and\ \bibinfo {author} {\bibfnamefont {F.~R.}\ \bibnamefont {Joaquim}},\
  }\href {\doibase 10.1103/RevModPhys.84.515} {\bibfield  {journal} {\bibinfo
  {journal} {Rev. Mod. Phys.}\ }\textbf {\bibinfo {volume} {84}},\ \bibinfo
  {pages} {515} (\bibinfo {year} {2012})},\ \Eprint
  {http://arxiv.org/abs/1111.5332} {arXiv:1111.5332 [hep-ph]} \BibitemShut
  {NoStop}%
\bibitem [{\citenamefont {Hagedorn}\ \emph {et~al.}(2018)\citenamefont
  {Hagedorn}, \citenamefont {Mohapatra}, \citenamefont {Molinaro},
  \citenamefont {Nishi},\ and\ \citenamefont {Petcov}}]{Hagedorn:2017wjy}%
  \BibitemOpen
  \bibfield  {author} {\bibinfo {author} {\bibfnamefont {C.}~\bibnamefont
  {Hagedorn}}, \bibinfo {author} {\bibfnamefont {R.~N.}\ \bibnamefont
  {Mohapatra}}, \bibinfo {author} {\bibfnamefont {E.}~\bibnamefont {Molinaro}},
  \bibinfo {author} {\bibfnamefont {C.~C.}\ \bibnamefont {Nishi}}, \ and\
  \bibinfo {author} {\bibfnamefont {S.~T.}\ \bibnamefont {Petcov}},\ }\href
  {\doibase 10.1142/S0217751X1842006X} {\bibfield  {journal} {\bibinfo
  {journal} {Int. J. Mod. Phys.}\ }\textbf {\bibinfo {volume} {A33}},\ \bibinfo
  {pages} {1842006} (\bibinfo {year} {2018})},\ \Eprint
  {http://arxiv.org/abs/1711.02866} {arXiv:1711.02866 [hep-ph]} \BibitemShut
  {NoStop}%
\bibitem [{\citenamefont {Grossman}(1995)}]{Grossman:1995wx}%
  \BibitemOpen
  \bibfield  {author} {\bibinfo {author} {\bibfnamefont {Y.}~\bibnamefont
  {Grossman}},\ }\href {\doibase 10.1016/0370-2693(95)01069-3} {\bibfield
  {journal} {\bibinfo  {journal} {Phys. Lett.}\ }\textbf {\bibinfo {volume}
  {B359}},\ \bibinfo {pages} {141} (\bibinfo {year} {1995})},\ \Eprint
  {http://arxiv.org/abs/hep-ph/9507344} {arXiv:hep-ph/9507344 [hep-ph]}
  \BibitemShut {NoStop}%
\bibitem [{\citenamefont {Friedland}\ \emph
  {et~al.}(2004{\natexlab{a}})\citenamefont {Friedland}, \citenamefont
  {Lunardini},\ and\ \citenamefont {Pena-Garay}}]{Friedland:2004pp}%
  \BibitemOpen
  \bibfield  {author} {\bibinfo {author} {\bibfnamefont {A.}~\bibnamefont
  {Friedland}}, \bibinfo {author} {\bibfnamefont {C.}~\bibnamefont
  {Lunardini}}, \ and\ \bibinfo {author} {\bibfnamefont {C.}~\bibnamefont
  {Pena-Garay}},\ }\href {\doibase 10.1016/j.physletb.2004.05.047} {\bibfield
  {journal} {\bibinfo  {journal} {Phys. Lett.}\ }\textbf {\bibinfo {volume}
  {B594}},\ \bibinfo {pages} {347} (\bibinfo {year} {2004}{\natexlab{a}})},\
  \Eprint {http://arxiv.org/abs/hep-ph/0402266} {arXiv:hep-ph/0402266 [hep-ph]}
  \BibitemShut {NoStop}%
\bibitem [{\citenamefont {Friedland}\ \emph
  {et~al.}(2004{\natexlab{b}})\citenamefont {Friedland}, \citenamefont
  {Lunardini},\ and\ \citenamefont {Maltoni}}]{Friedland:2004ah}%
  \BibitemOpen
  \bibfield  {author} {\bibinfo {author} {\bibfnamefont {A.}~\bibnamefont
  {Friedland}}, \bibinfo {author} {\bibfnamefont {C.}~\bibnamefont
  {Lunardini}}, \ and\ \bibinfo {author} {\bibfnamefont {M.}~\bibnamefont
  {Maltoni}},\ }\href {\doibase 10.1103/PhysRevD.70.111301} {\bibfield
  {journal} {\bibinfo  {journal} {Phys. Rev.}\ }\textbf {\bibinfo {volume}
  {D70}},\ \bibinfo {pages} {111301} (\bibinfo {year} {2004}{\natexlab{b}})},\
  \Eprint {http://arxiv.org/abs/hep-ph/0408264} {arXiv:hep-ph/0408264 [hep-ph]}
  \BibitemShut {NoStop}%
\bibitem [{\citenamefont {Antusch}\ \emph {et~al.}(2009)\citenamefont
  {Antusch}, \citenamefont {Baumann},\ and\ \citenamefont
  {Fernandez-Martinez}}]{Antusch:2008tz}%
  \BibitemOpen
  \bibfield  {author} {\bibinfo {author} {\bibfnamefont {S.}~\bibnamefont
  {Antusch}}, \bibinfo {author} {\bibfnamefont {J.~P.}\ \bibnamefont
  {Baumann}}, \ and\ \bibinfo {author} {\bibfnamefont {E.}~\bibnamefont
  {Fernandez-Martinez}},\ }\href {\doibase 10.1016/j.nuclphysb.2008.11.018}
  {\bibfield  {journal} {\bibinfo  {journal} {Nucl. Phys.}\ }\textbf {\bibinfo
  {volume} {B810}},\ \bibinfo {pages} {369} (\bibinfo {year} {2009})},\ \Eprint
  {http://arxiv.org/abs/0807.1003} {arXiv:0807.1003 [hep-ph]} \BibitemShut
  {NoStop}%
\bibitem [{\citenamefont {Miranda}\ and\ \citenamefont
  {Nunokawa}(2015)}]{Miranda:2015dra}%
  \BibitemOpen
  \bibfield  {author} {\bibinfo {author} {\bibfnamefont {O.~G.}\ \bibnamefont
  {Miranda}}\ and\ \bibinfo {author} {\bibfnamefont {H.}~\bibnamefont
  {Nunokawa}},\ }\href {\doibase 10.1088/1367-2630/17/9/095002} {\bibfield
  {journal} {\bibinfo  {journal} {New J. Phys.}\ }\textbf {\bibinfo {volume}
  {17}},\ \bibinfo {pages} {095002} (\bibinfo {year} {2015})},\ \Eprint
  {http://arxiv.org/abs/1505.06254} {arXiv:1505.06254 [hep-ph]} \BibitemShut
  {NoStop}%
\bibitem [{\citenamefont {Farzan}\ and\ \citenamefont
  {Shoemaker}(2016)}]{Farzan:2015hkd}%
  \BibitemOpen
  \bibfield  {author} {\bibinfo {author} {\bibfnamefont {Y.}~\bibnamefont
  {Farzan}}\ and\ \bibinfo {author} {\bibfnamefont {I.~M.}\ \bibnamefont
  {Shoemaker}},\ }\href {\doibase 10.1007/JHEP07(2016)033} {\bibfield
  {journal} {\bibinfo  {journal} {JHEP}\ }\textbf {\bibinfo {volume} {07}},\
  \bibinfo {pages} {033} (\bibinfo {year} {2016})},\ \Eprint
  {http://arxiv.org/abs/1512.09147} {arXiv:1512.09147 [hep-ph]} \BibitemShut
  {NoStop}%
\bibitem [{\citenamefont {Arafune}\ \emph {et~al.}(1997)\citenamefont
  {Arafune}, \citenamefont {Koike},\ and\ \citenamefont
  {Sato}}]{Arafune:1997hd}%
  \BibitemOpen
  \bibfield  {author} {\bibinfo {author} {\bibfnamefont {J.}~\bibnamefont
  {Arafune}}, \bibinfo {author} {\bibfnamefont {M.}~\bibnamefont {Koike}}, \
  and\ \bibinfo {author} {\bibfnamefont {J.}~\bibnamefont {Sato}},\ }\href
  {\doibase 10.1103/PhysRevD.60.119905, 10.1103/PhysRevD.56.3093} {\bibfield
  {journal} {\bibinfo  {journal} {Phys. Rev.}\ }\textbf {\bibinfo {volume}
  {D56}},\ \bibinfo {pages} {3093} (\bibinfo {year} {1997})},\ \bibinfo {note}
  {[Erratum: Phys. Rev.D60,119905(1999)]},\ \Eprint
  {http://arxiv.org/abs/hep-ph/9703351} {arXiv:hep-ph/9703351 [hep-ph]}
  \BibitemShut {NoStop}%
\bibitem [{\citenamefont {Barger}\ \emph {et~al.}(2002)\citenamefont {Barger},
  \citenamefont {Marfatia},\ and\ \citenamefont {Whisnant}}]{Barger:2001yr}%
  \BibitemOpen
  \bibfield  {author} {\bibinfo {author} {\bibfnamefont {V.}~\bibnamefont
  {Barger}}, \bibinfo {author} {\bibfnamefont {D.}~\bibnamefont {Marfatia}}, \
  and\ \bibinfo {author} {\bibfnamefont {K.}~\bibnamefont {Whisnant}},\ }\href
  {\doibase 10.1103/PhysRevD.65.073023} {\bibfield  {journal} {\bibinfo
  {journal} {Phys. Rev.}\ }\textbf {\bibinfo {volume} {D65}},\ \bibinfo {pages}
  {073023} (\bibinfo {year} {2002})},\ \Eprint
  {http://arxiv.org/abs/hep-ph/0112119} {arXiv:hep-ph/0112119 [hep-ph]}
  \BibitemShut {NoStop}%
\bibitem [{\citenamefont {Minakata}\ and\ \citenamefont
  {Nunokawa}(2001)}]{Minakata:2001qm}%
  \BibitemOpen
  \bibfield  {author} {\bibinfo {author} {\bibfnamefont {H.}~\bibnamefont
  {Minakata}}\ and\ \bibinfo {author} {\bibfnamefont {H.}~\bibnamefont
  {Nunokawa}},\ }\href {\doibase 10.1088/1126-6708/2001/10/001} {\bibfield
  {journal} {\bibinfo  {journal} {JHEP}\ }\textbf {\bibinfo {volume} {10}},\
  \bibinfo {pages} {001} (\bibinfo {year} {2001})},\ \Eprint
  {http://arxiv.org/abs/hep-ph/0108085} {arXiv:hep-ph/0108085 [hep-ph]}
  \BibitemShut {NoStop}%
\bibitem [{\citenamefont {Ohlsson}\ \emph {et~al.}(2013)\citenamefont
  {Ohlsson}, \citenamefont {Zhang},\ and\ \citenamefont
  {Zhou}}]{Ohlsson:2013ip}%
  \BibitemOpen
  \bibfield  {author} {\bibinfo {author} {\bibfnamefont {T.}~\bibnamefont
  {Ohlsson}}, \bibinfo {author} {\bibfnamefont {H.}~\bibnamefont {Zhang}}, \
  and\ \bibinfo {author} {\bibfnamefont {S.}~\bibnamefont {Zhou}},\ }\href
  {\doibase 10.1103/PhysRevD.87.053006} {\bibfield  {journal} {\bibinfo
  {journal} {Phys. Rev.}\ }\textbf {\bibinfo {volume} {D87}},\ \bibinfo {pages}
  {053006} (\bibinfo {year} {2013})},\ \Eprint {http://arxiv.org/abs/1301.4333}
  {arXiv:1301.4333 [hep-ph]} \BibitemShut {NoStop}%
\bibitem [{\citenamefont {Gonzalez-Garcia}\ \emph {et~al.}(2001)\citenamefont
  {Gonzalez-Garcia}, \citenamefont {Grossman}, \citenamefont {Gusso},\ and\
  \citenamefont {Nir}}]{GonzalezGarcia:2001mp}%
  \BibitemOpen
  \bibfield  {author} {\bibinfo {author} {\bibfnamefont {M.~C.}\ \bibnamefont
  {Gonzalez-Garcia}}, \bibinfo {author} {\bibfnamefont {Y.}~\bibnamefont
  {Grossman}}, \bibinfo {author} {\bibfnamefont {A.}~\bibnamefont {Gusso}}, \
  and\ \bibinfo {author} {\bibfnamefont {Y.}~\bibnamefont {Nir}},\ }\href
  {\doibase 10.1103/PhysRevD.64.096006} {\bibfield  {journal} {\bibinfo
  {journal} {Phys. Rev.}\ }\textbf {\bibinfo {volume} {D64}},\ \bibinfo {pages}
  {096006} (\bibinfo {year} {2001})},\ \Eprint
  {http://arxiv.org/abs/hep-ph/0105159} {arXiv:hep-ph/0105159 [hep-ph]}
  \BibitemShut {NoStop}%
\bibitem [{\citenamefont {Coloma}\ \emph {et~al.}(2011)\citenamefont {Coloma},
  \citenamefont {Donini}, \citenamefont {Lopez-Pavon},\ and\ \citenamefont
  {Minakata}}]{Coloma:2011rq}%
  \BibitemOpen
  \bibfield  {author} {\bibinfo {author} {\bibfnamefont {P.}~\bibnamefont
  {Coloma}}, \bibinfo {author} {\bibfnamefont {A.}~\bibnamefont {Donini}},
  \bibinfo {author} {\bibfnamefont {J.}~\bibnamefont {Lopez-Pavon}}, \ and\
  \bibinfo {author} {\bibfnamefont {H.}~\bibnamefont {Minakata}},\ }\href
  {\doibase 10.1007/JHEP08(2011)036} {\bibfield  {journal} {\bibinfo  {journal}
  {JHEP}\ }\textbf {\bibinfo {volume} {08}},\ \bibinfo {pages} {036} (\bibinfo
  {year} {2011})},\ \Eprint {http://arxiv.org/abs/1105.5936} {arXiv:1105.5936
  [hep-ph]} \BibitemShut {NoStop}%
\bibitem [{\citenamefont {Friedland}\ and\ \citenamefont
  {Shoemaker}(2012)}]{Friedland:2012tq}%
  \BibitemOpen
  \bibfield  {author} {\bibinfo {author} {\bibfnamefont {A.}~\bibnamefont
  {Friedland}}\ and\ \bibinfo {author} {\bibfnamefont {I.~M.}\ \bibnamefont
  {Shoemaker}},\ }\href@noop {} {\  (\bibinfo {year} {2012})},\ \Eprint
  {http://arxiv.org/abs/1207.6642} {arXiv:1207.6642 [hep-ph]} \BibitemShut
  {NoStop}%
\bibitem [{\citenamefont {Rahman}\ \emph {et~al.}(2015)\citenamefont {Rahman},
  \citenamefont {Dasgupta},\ and\ \citenamefont {Adhikari}}]{Rahman:2015vqa}%
  \BibitemOpen
  \bibfield  {author} {\bibinfo {author} {\bibfnamefont {Z.}~\bibnamefont
  {Rahman}}, \bibinfo {author} {\bibfnamefont {A.}~\bibnamefont {Dasgupta}}, \
  and\ \bibinfo {author} {\bibfnamefont {R.}~\bibnamefont {Adhikari}},\ }\href
  {\doibase 10.1088/0954-3899/42/6/065001} {\bibfield  {journal} {\bibinfo
  {journal} {J. Phys.}\ }\textbf {\bibinfo {volume} {G42}},\ \bibinfo {pages}
  {065001} (\bibinfo {year} {2015})},\ \Eprint
  {http://arxiv.org/abs/1503.03248} {arXiv:1503.03248 [hep-ph]} \BibitemShut
  {NoStop}%
\bibitem [{\citenamefont {Masud}\ \emph {et~al.}(2016)\citenamefont {Masud},
  \citenamefont {Chatterjee},\ and\ \citenamefont {Mehta}}]{Masud:2015xva}%
  \BibitemOpen
  \bibfield  {author} {\bibinfo {author} {\bibfnamefont {M.}~\bibnamefont
  {Masud}}, \bibinfo {author} {\bibfnamefont {A.}~\bibnamefont {Chatterjee}}, \
  and\ \bibinfo {author} {\bibfnamefont {P.}~\bibnamefont {Mehta}},\ }\href
  {\doibase 10.1088/0954-3899/43/9/095005/meta, 10.1088/0954-3899/43/9/095005}
  {\bibfield  {journal} {\bibinfo  {journal} {J. Phys.}\ }\textbf {\bibinfo
  {volume} {G43}},\ \bibinfo {pages} {095005} (\bibinfo {year} {2016})},\
  \Eprint {http://arxiv.org/abs/1510.08261} {arXiv:1510.08261 [hep-ph]}
  \BibitemShut {NoStop}%
\bibitem [{\citenamefont {Coloma}(2016)}]{Coloma:2015kiu}%
  \BibitemOpen
  \bibfield  {author} {\bibinfo {author} {\bibfnamefont {P.}~\bibnamefont
  {Coloma}},\ }\href {\doibase 10.1007/JHEP03(2016)016} {\bibfield  {journal}
  {\bibinfo  {journal} {JHEP}\ }\textbf {\bibinfo {volume} {03}},\ \bibinfo
  {pages} {016} (\bibinfo {year} {2016})},\ \Eprint
  {http://arxiv.org/abs/1511.06357} {arXiv:1511.06357 [hep-ph]} \BibitemShut
  {NoStop}%
\bibitem [{\citenamefont {Palazzo}(2016)}]{Palazzo:2015gja}%
  \BibitemOpen
  \bibfield  {author} {\bibinfo {author} {\bibfnamefont {A.}~\bibnamefont
  {Palazzo}},\ }\href {\doibase 10.1016/j.physletb.2016.03.061} {\bibfield
  {journal} {\bibinfo  {journal} {Phys. Lett.}\ }\textbf {\bibinfo {volume}
  {B757}},\ \bibinfo {pages} {142} (\bibinfo {year} {2016})},\ \Eprint
  {http://arxiv.org/abs/1509.03148} {arXiv:1509.03148 [hep-ph]} \BibitemShut
  {NoStop}%
\bibitem [{\citenamefont {de~Gouvêa}\ and\ \citenamefont
  {Kelly}(2016{\natexlab{a}})}]{deGouvea:2015ndi}%
  \BibitemOpen
  \bibfield  {author} {\bibinfo {author} {\bibfnamefont {A.}~\bibnamefont
  {de~Gouvêa}}\ and\ \bibinfo {author} {\bibfnamefont {K.~J.}\ \bibnamefont
  {Kelly}},\ }\href {\doibase 10.1016/j.nuclphysb.2016.03.013} {\bibfield
  {journal} {\bibinfo  {journal} {Nucl. Phys.}\ }\textbf {\bibinfo {volume}
  {B908}},\ \bibinfo {pages} {318} (\bibinfo {year} {2016}{\natexlab{a}})},\
  \Eprint {http://arxiv.org/abs/1511.05562} {arXiv:1511.05562 [hep-ph]}
  \BibitemShut {NoStop}%
\bibitem [{\citenamefont {Masud}\ and\ \citenamefont
  {Mehta}(2016)}]{Masud:2016bvp}%
  \BibitemOpen
  \bibfield  {author} {\bibinfo {author} {\bibfnamefont {M.}~\bibnamefont
  {Masud}}\ and\ \bibinfo {author} {\bibfnamefont {P.}~\bibnamefont {Mehta}},\
  }\href {\doibase 10.1103/PhysRevD.94.013014} {\bibfield  {journal} {\bibinfo
  {journal} {Phys. Rev.}\ }\textbf {\bibinfo {volume} {D94}},\ \bibinfo {pages}
  {013014} (\bibinfo {year} {2016})},\ \Eprint
  {http://arxiv.org/abs/1603.01380} {arXiv:1603.01380 [hep-ph]} \BibitemShut
  {NoStop}%
\bibitem [{\citenamefont {Blennow}\ \emph {et~al.}(2016)\citenamefont
  {Blennow}, \citenamefont {Choubey}, \citenamefont {Ohlsson}, \citenamefont
  {Pramanik},\ and\ \citenamefont {Raut}}]{Blennow:2016etl}%
  \BibitemOpen
  \bibfield  {author} {\bibinfo {author} {\bibfnamefont {M.}~\bibnamefont
  {Blennow}}, \bibinfo {author} {\bibfnamefont {S.}~\bibnamefont {Choubey}},
  \bibinfo {author} {\bibfnamefont {T.}~\bibnamefont {Ohlsson}}, \bibinfo
  {author} {\bibfnamefont {D.}~\bibnamefont {Pramanik}}, \ and\ \bibinfo
  {author} {\bibfnamefont {S.~K.}\ \bibnamefont {Raut}},\ }\href {\doibase
  10.1007/JHEP08(2016)090} {\bibfield  {journal} {\bibinfo  {journal} {JHEP}\
  }\textbf {\bibinfo {volume} {08}},\ \bibinfo {pages} {090} (\bibinfo {year}
  {2016})},\ \Eprint {http://arxiv.org/abs/1606.08851} {arXiv:1606.08851
  [hep-ph]} \BibitemShut {NoStop}%
\bibitem [{\citenamefont {Liao}\ \emph {et~al.}(2016)\citenamefont {Liao},
  \citenamefont {Marfatia},\ and\ \citenamefont {Whisnant}}]{Liao:2016hsa}%
  \BibitemOpen
  \bibfield  {author} {\bibinfo {author} {\bibfnamefont {J.}~\bibnamefont
  {Liao}}, \bibinfo {author} {\bibfnamefont {D.}~\bibnamefont {Marfatia}}, \
  and\ \bibinfo {author} {\bibfnamefont {K.}~\bibnamefont {Whisnant}},\ }\href
  {\doibase 10.1103/PhysRevD.93.093016} {\bibfield  {journal} {\bibinfo
  {journal} {Phys. Rev.}\ }\textbf {\bibinfo {volume} {D93}},\ \bibinfo {pages}
  {093016} (\bibinfo {year} {2016})},\ \Eprint
  {http://arxiv.org/abs/1601.00927} {arXiv:1601.00927 [hep-ph]} \BibitemShut
  {NoStop}%
\bibitem [{\citenamefont {de~Gouvêa}\ and\ \citenamefont
  {Kelly}(2016{\natexlab{b}})}]{deGouvea:2016pom}%
  \BibitemOpen
  \bibfield  {author} {\bibinfo {author} {\bibfnamefont {A.}~\bibnamefont
  {de~Gouvêa}}\ and\ \bibinfo {author} {\bibfnamefont {K.~J.}\ \bibnamefont
  {Kelly}},\ }\href@noop {} {\  (\bibinfo {year} {2016}{\natexlab{b}})},\
  \Eprint {http://arxiv.org/abs/1605.09376} {arXiv:1605.09376 [hep-ph]}
  \BibitemShut {NoStop}%
\bibitem [{\citenamefont {Liao}\ \emph {et~al.}(2017)\citenamefont {Liao},
  \citenamefont {Marfatia},\ and\ \citenamefont {Whisnant}}]{Liao:2016orc}%
  \BibitemOpen
  \bibfield  {author} {\bibinfo {author} {\bibfnamefont {J.}~\bibnamefont
  {Liao}}, \bibinfo {author} {\bibfnamefont {D.}~\bibnamefont {Marfatia}}, \
  and\ \bibinfo {author} {\bibfnamefont {K.}~\bibnamefont {Whisnant}},\ }\href
  {\doibase 10.1007/JHEP01(2017)071} {\bibfield  {journal} {\bibinfo  {journal}
  {JHEP}\ }\textbf {\bibinfo {volume} {01}},\ \bibinfo {pages} {071} (\bibinfo
  {year} {2017})},\ \Eprint {http://arxiv.org/abs/1612.01443} {arXiv:1612.01443
  [hep-ph]} \BibitemShut {NoStop}%
\bibitem [{\citenamefont {Ge}\ and\ \citenamefont
  {Smirnov}(2016)}]{Ge:2016dlx}%
  \BibitemOpen
  \bibfield  {author} {\bibinfo {author} {\bibfnamefont {S.-F.}\ \bibnamefont
  {Ge}}\ and\ \bibinfo {author} {\bibfnamefont {A.~{\relax Yu}.}\ \bibnamefont
  {Smirnov}},\ }\href {\doibase 10.1007/JHEP10(2016)138} {\bibfield  {journal}
  {\bibinfo  {journal} {JHEP}\ }\textbf {\bibinfo {volume} {10}},\ \bibinfo
  {pages} {138} (\bibinfo {year} {2016})},\ \Eprint
  {http://arxiv.org/abs/1607.08513} {arXiv:1607.08513 [hep-ph]} \BibitemShut
  {NoStop}%
\bibitem [{\citenamefont {Agarwalla}\ \emph {et~al.}(2016)\citenamefont
  {Agarwalla}, \citenamefont {Chatterjee},\ and\ \citenamefont
  {Palazzo}}]{Agarwalla:2016fkh}%
  \BibitemOpen
  \bibfield  {author} {\bibinfo {author} {\bibfnamefont {S.~K.}\ \bibnamefont
  {Agarwalla}}, \bibinfo {author} {\bibfnamefont {S.~S.}\ \bibnamefont
  {Chatterjee}}, \ and\ \bibinfo {author} {\bibfnamefont {A.}~\bibnamefont
  {Palazzo}},\ }\href {\doibase 10.1016/j.physletb.2016.09.020} {\bibfield
  {journal} {\bibinfo  {journal} {Phys. Lett.}\ }\textbf {\bibinfo {volume}
  {B762}},\ \bibinfo {pages} {64} (\bibinfo {year} {2016})},\ \Eprint
  {http://arxiv.org/abs/1607.01745} {arXiv:1607.01745 [hep-ph]} \BibitemShut
  {NoStop}%
\bibitem [{\citenamefont {Fukasawa}\ \emph {et~al.}(2017)\citenamefont
  {Fukasawa}, \citenamefont {Ghosh},\ and\ \citenamefont
  {Yasuda}}]{Fukasawa:2016lew}%
  \BibitemOpen
  \bibfield  {author} {\bibinfo {author} {\bibfnamefont {S.}~\bibnamefont
  {Fukasawa}}, \bibinfo {author} {\bibfnamefont {M.}~\bibnamefont {Ghosh}}, \
  and\ \bibinfo {author} {\bibfnamefont {O.}~\bibnamefont {Yasuda}},\ }\href
  {\doibase 10.1103/PhysRevD.95.055005} {\bibfield  {journal} {\bibinfo
  {journal} {Phys. Rev.}\ }\textbf {\bibinfo {volume} {D95}},\ \bibinfo {pages}
  {055005} (\bibinfo {year} {2017})},\ \Eprint
  {http://arxiv.org/abs/1611.06141} {arXiv:1611.06141 [hep-ph]} \BibitemShut
  {NoStop}%
\bibitem [{\citenamefont {Deepthi}\ \emph
  {et~al.}(2017{\natexlab{a}})\citenamefont {Deepthi}, \citenamefont
  {Goswami},\ and\ \citenamefont {Nath}}]{Deepthi:2016erc}%
  \BibitemOpen
  \bibfield  {author} {\bibinfo {author} {\bibfnamefont {K.~N.}\ \bibnamefont
  {Deepthi}}, \bibinfo {author} {\bibfnamefont {S.}~\bibnamefont {Goswami}}, \
  and\ \bibinfo {author} {\bibfnamefont {N.}~\bibnamefont {Nath}},\ }\href
  {\doibase 10.1103/PhysRevD.96.075023} {\bibfield  {journal} {\bibinfo
  {journal} {Phys. Rev.}\ }\textbf {\bibinfo {volume} {D96}},\ \bibinfo {pages}
  {075023} (\bibinfo {year} {2017}{\natexlab{a}})},\ \Eprint
  {http://arxiv.org/abs/1612.00784} {arXiv:1612.00784 [hep-ph]} \BibitemShut
  {NoStop}%
\bibitem [{\citenamefont {de~Gouvêa}\ and\ \citenamefont
  {Kelly}(2017)}]{deGouvea:2017yvn}%
  \BibitemOpen
  \bibfield  {author} {\bibinfo {author} {\bibfnamefont {A.}~\bibnamefont
  {de~Gouvêa}}\ and\ \bibinfo {author} {\bibfnamefont {K.~J.}\ \bibnamefont
  {Kelly}},\ }\href {\doibase 10.1103/PhysRevD.96.095018} {\bibfield  {journal}
  {\bibinfo  {journal} {Phys. Rev.}\ }\textbf {\bibinfo {volume} {D96}},\
  \bibinfo {pages} {095018} (\bibinfo {year} {2017})},\ \Eprint
  {http://arxiv.org/abs/1709.06090} {arXiv:1709.06090 [hep-ph]} \BibitemShut
  {NoStop}%
\bibitem [{\citenamefont {Deepthi}\ \emph
  {et~al.}(2017{\natexlab{b}})\citenamefont {Deepthi}, \citenamefont
  {Goswami},\ and\ \citenamefont {Nath}}]{Deepthi:2017gxg}%
  \BibitemOpen
  \bibfield  {author} {\bibinfo {author} {\bibfnamefont {K.~N.}\ \bibnamefont
  {Deepthi}}, \bibinfo {author} {\bibfnamefont {S.}~\bibnamefont {Goswami}}, \
  and\ \bibinfo {author} {\bibfnamefont {N.}~\bibnamefont {Nath}},\ }\href@noop
  {} {\  (\bibinfo {year} {2017}{\natexlab{b}})},\ \Eprint
  {http://arxiv.org/abs/1711.04840} {arXiv:1711.04840 [hep-ph]} \BibitemShut
  {NoStop}%
\bibitem [{\citenamefont {Ballett}\ \emph {et~al.}(2017)\citenamefont
  {Ballett}, \citenamefont {King}, \citenamefont {Pascoli}, \citenamefont
  {Prouse},\ and\ \citenamefont {Wang}}]{Ballett:2016daj}%
  \BibitemOpen
  \bibfield  {author} {\bibinfo {author} {\bibfnamefont {P.}~\bibnamefont
  {Ballett}}, \bibinfo {author} {\bibfnamefont {S.~F.}\ \bibnamefont {King}},
  \bibinfo {author} {\bibfnamefont {S.}~\bibnamefont {Pascoli}}, \bibinfo
  {author} {\bibfnamefont {N.~W.}\ \bibnamefont {Prouse}}, \ and\ \bibinfo
  {author} {\bibfnamefont {T.}~\bibnamefont {Wang}},\ }\href {\doibase
  10.1103/PhysRevD.96.033003} {\bibfield  {journal} {\bibinfo  {journal} {Phys.
  Rev.}\ }\textbf {\bibinfo {volume} {D96}},\ \bibinfo {pages} {033003}
  (\bibinfo {year} {2017})},\ \Eprint {http://arxiv.org/abs/1612.07275}
  {arXiv:1612.07275 [hep-ph]} \BibitemShut {NoStop}%
\bibitem [{\citenamefont {Abe}\ \emph {et~al.}(2017)\citenamefont {Abe} \emph
  {et~al.}}]{Abe:2017vif}%
  \BibitemOpen
  \bibfield  {author} {\bibinfo {author} {\bibfnamefont {K.}~\bibnamefont
  {Abe}} \emph {et~al.} (\bibinfo {collaboration} {T2K}),\ }\href {\doibase
  10.1103/PhysRevD.96.092006} {\bibfield  {journal} {\bibinfo  {journal} {Phys.
  Rev.}\ }\textbf {\bibinfo {volume} {D96}},\ \bibinfo {pages} {092006}
  (\bibinfo {year} {2017})},\ \Eprint {http://arxiv.org/abs/1707.01048}
  {arXiv:1707.01048 [hep-ex]} \BibitemShut {NoStop}%
\bibitem [{\citenamefont {Acero}\ \emph {et~al.}(2018)\citenamefont {Acero}
  \emph {et~al.}}]{NOvA:2018gge}%
  \BibitemOpen
  \bibfield  {author} {\bibinfo {author} {\bibfnamefont {M.~A.}\ \bibnamefont
  {Acero}} \emph {et~al.} (\bibinfo {collaboration} {NOvA}),\ }\href@noop {} {\
   (\bibinfo {year} {2018})},\ \Eprint {http://arxiv.org/abs/1806.00096}
  {arXiv:1806.00096 [hep-ex]} \BibitemShut {NoStop}%
\bibitem [{\citenamefont {Gago}\ \emph {et~al.}(2010)\citenamefont {Gago},
  \citenamefont {Minakata}, \citenamefont {Nunokawa}, \citenamefont
  {Uchinami},\ and\ \citenamefont {Zukanovich~Funchal}}]{Gago:2009ij}%
  \BibitemOpen
  \bibfield  {author} {\bibinfo {author} {\bibfnamefont {A.~M.}\ \bibnamefont
  {Gago}}, \bibinfo {author} {\bibfnamefont {H.}~\bibnamefont {Minakata}},
  \bibinfo {author} {\bibfnamefont {H.}~\bibnamefont {Nunokawa}}, \bibinfo
  {author} {\bibfnamefont {S.}~\bibnamefont {Uchinami}}, \ and\ \bibinfo
  {author} {\bibfnamefont {R.}~\bibnamefont {Zukanovich~Funchal}},\ }\href
  {\doibase 10.1007/JHEP01(2010)049} {\bibfield  {journal} {\bibinfo  {journal}
  {JHEP}\ }\textbf {\bibinfo {volume} {01}},\ \bibinfo {pages} {049} (\bibinfo
  {year} {2010})},\ \Eprint {http://arxiv.org/abs/0904.3360} {arXiv:0904.3360
  [hep-ph]} \BibitemShut {NoStop}%
\bibitem [{\citenamefont {Soumya}\ and\ \citenamefont
  {Mohanta}(2016)}]{Soumya:2016enw}%
  \BibitemOpen
  \bibfield  {author} {\bibinfo {author} {\bibfnamefont {C.}~\bibnamefont
  {Soumya}}\ and\ \bibinfo {author} {\bibfnamefont {R.}~\bibnamefont
  {Mohanta}},\ }\href@noop {} {\  (\bibinfo {year} {2016})},\ \Eprint
  {http://arxiv.org/abs/1603.02184} {arXiv:1603.02184 [hep-ph]} \BibitemShut
  {NoStop}%
\bibitem [{\citenamefont {Flores}\ \emph {et~al.}(2018)\citenamefont {Flores},
  \citenamefont {Garcés},\ and\ \citenamefont {Miranda}}]{Flores:2018kwk}%
  \BibitemOpen
  \bibfield  {author} {\bibinfo {author} {\bibfnamefont {L.~J.}\ \bibnamefont
  {Flores}}, \bibinfo {author} {\bibfnamefont {E.~A.}\ \bibnamefont {Garcés}},
  \ and\ \bibinfo {author} {\bibfnamefont {O.~G.}\ \bibnamefont {Miranda}},\
  }\href@noop {} {\  (\bibinfo {year} {2018})},\ \Eprint
  {http://arxiv.org/abs/1806.07951} {arXiv:1806.07951 [hep-ph]} \BibitemShut
  {NoStop}%
\bibitem [{\citenamefont {Patrignani}\ \emph {et~al.}(2016)\citenamefont
  {Patrignani} \emph {et~al.}}]{Patrignani:2016xqp}%
  \BibitemOpen
  \bibfield  {author} {\bibinfo {author} {\bibfnamefont {C.}~\bibnamefont
  {Patrignani}} \emph {et~al.} (\bibinfo {collaboration} {Particle Data
  Group}),\ }\href {\doibase 10.1088/1674-1137/40/10/100001} {\bibfield
  {journal} {\bibinfo  {journal} {Chin. Phys.}\ }\textbf {\bibinfo {volume}
  {C40}},\ \bibinfo {pages} {100001} (\bibinfo {year} {2016})}\BibitemShut
  {NoStop}%
\bibitem [{\citenamefont {Ohlsson}(2013)}]{Ohlsson:2012kf}%
  \BibitemOpen
  \bibfield  {author} {\bibinfo {author} {\bibfnamefont {T.}~\bibnamefont
  {Ohlsson}},\ }\href {\doibase 10.1088/0034-4885/76/4/044201} {\bibfield
  {journal} {\bibinfo  {journal} {Rept. Prog. Phys.}\ }\textbf {\bibinfo
  {volume} {76}},\ \bibinfo {pages} {044201} (\bibinfo {year} {2013})},\
  \Eprint {http://arxiv.org/abs/1209.2710} {arXiv:1209.2710 [hep-ph]}
  \BibitemShut {NoStop}%
\bibitem [{\citenamefont {Cervera}\ \emph {et~al.}(2000)\citenamefont
  {Cervera}, \citenamefont {Donini}, \citenamefont {Gavela}, \citenamefont
  {Gomez~Cadenas}, \citenamefont {Hernandez}, \citenamefont {Mena},\ and\
  \citenamefont {Rigolin}}]{Cervera:2000kp}%
  \BibitemOpen
  \bibfield  {author} {\bibinfo {author} {\bibfnamefont {A.}~\bibnamefont
  {Cervera}}, \bibinfo {author} {\bibfnamefont {A.}~\bibnamefont {Donini}},
  \bibinfo {author} {\bibfnamefont {M.~B.}\ \bibnamefont {Gavela}}, \bibinfo
  {author} {\bibfnamefont {J.~J.}\ \bibnamefont {Gomez~Cadenas}}, \bibinfo
  {author} {\bibfnamefont {P.}~\bibnamefont {Hernandez}}, \bibinfo {author}
  {\bibfnamefont {O.}~\bibnamefont {Mena}}, \ and\ \bibinfo {author}
  {\bibfnamefont {S.}~\bibnamefont {Rigolin}},\ }\href {\doibase
  10.1016/S0550-3213(00)00606-4, 10.1016/S0550-3213(00)00221-2} {\bibfield
  {journal} {\bibinfo  {journal} {Nucl. Phys.}\ }\textbf {\bibinfo {volume}
  {B579}},\ \bibinfo {pages} {17} (\bibinfo {year} {2000})},\ \bibinfo {note}
  {[Erratum: Nucl. Phys.B593,731(2001)]},\ \Eprint
  {http://arxiv.org/abs/hep-ph/0002108} {arXiv:hep-ph/0002108 [hep-ph]}
  \BibitemShut {NoStop}%
\bibitem [{\citenamefont {Kikuchi}\ \emph {et~al.}(2009)\citenamefont
  {Kikuchi}, \citenamefont {Minakata},\ and\ \citenamefont
  {Uchinami}}]{Kikuchi:2008vq}%
  \BibitemOpen
  \bibfield  {author} {\bibinfo {author} {\bibfnamefont {T.}~\bibnamefont
  {Kikuchi}}, \bibinfo {author} {\bibfnamefont {H.}~\bibnamefont {Minakata}}, \
  and\ \bibinfo {author} {\bibfnamefont {S.}~\bibnamefont {Uchinami}},\ }\href
  {\doibase 10.1088/1126-6708/2009/03/114} {\bibfield  {journal} {\bibinfo
  {journal} {JHEP}\ }\textbf {\bibinfo {volume} {03}},\ \bibinfo {pages} {114}
  (\bibinfo {year} {2009})},\ \Eprint {http://arxiv.org/abs/0809.3312}
  {arXiv:0809.3312 [hep-ph]} \BibitemShut {NoStop}%
\bibitem [{\citenamefont {Forero}\ and\ \citenamefont
  {Huber}(2016)}]{Forero:2016cmb}%
  \BibitemOpen
  \bibfield  {author} {\bibinfo {author} {\bibfnamefont {D.~V.}\ \bibnamefont
  {Forero}}\ and\ \bibinfo {author} {\bibfnamefont {P.}~\bibnamefont {Huber}},\
  }\href {\doibase 10.1103/PhysRevLett.117.031801} {\bibfield  {journal}
  {\bibinfo  {journal} {Phys. Rev. Lett.}\ }\textbf {\bibinfo {volume} {117}},\
  \bibinfo {pages} {031801} (\bibinfo {year} {2016})},\ \Eprint
  {http://arxiv.org/abs/1601.03736} {arXiv:1601.03736 [hep-ph]} \BibitemShut
  {NoStop}%
\bibitem [{\citenamefont {Abe}\ \emph {et~al.}(2018)\citenamefont {Abe} \emph
  {et~al.}}]{Abe:2016ero}%
  \BibitemOpen
  \bibfield  {author} {\bibinfo {author} {\bibfnamefont {K.}~\bibnamefont
  {Abe}} \emph {et~al.} (\bibinfo {collaboration} {Hyper-Kamiokande}),\ }\href
  {\doibase 10.1093/ptep/pty044} {\bibfield  {journal} {\bibinfo  {journal}
  {PTEP}\ }\textbf {\bibinfo {volume} {2018}},\ \bibinfo {pages} {063C01}
  (\bibinfo {year} {2018})},\ \Eprint {http://arxiv.org/abs/1611.06118}
  {arXiv:1611.06118 [hep-ex]} \BibitemShut {NoStop}%
\bibitem [{\citenamefont {Beavis}\ \emph {et~al.}(1995)\citenamefont {Beavis}
  \emph {et~al.}}]{Beavis:1995pbs}%
  \BibitemOpen
  \bibfield  {author} {\bibinfo {author} {\bibfnamefont {D.}~\bibnamefont
  {Beavis}} \emph {et~al.} (\bibinfo {collaboration} {E899}),\ }\href {\doibase
  10.2172/52878} {\  (\bibinfo {year} {1995}),\ 10.2172/52878}\BibitemShut
  {NoStop}%
\bibitem [{\citenamefont {Itow}\ \emph {et~al.}(2001)\citenamefont {Itow} \emph
  {et~al.}}]{Itow:2001ee}%
  \BibitemOpen
  \bibfield  {author} {\bibinfo {author} {\bibfnamefont {Y.}~\bibnamefont
  {Itow}} \emph {et~al.} (\bibinfo {collaboration} {T2K}),\ }in\ \href@noop {}
  {\emph {\bibinfo {booktitle} {{Neutrino oscillations and their origin.
  Proceedings, 3rd International Workshop, NOON 2001, Kashiwa, Tokyo, Japan,
  December 508, 2001}}}}\ (\bibinfo {year} {2001})\ pp.\ \bibinfo {pages}
  {239--248},\ \Eprint {http://arxiv.org/abs/hep-ex/0106019}
  {arXiv:hep-ex/0106019 [hep-ex]} \BibitemShut {NoStop}%
\bibitem [{\citenamefont {Ayres}\ \emph {et~al.}(2002)\citenamefont {Ayres}
  \emph {et~al.}}]{Ayres:2002ws}%
  \BibitemOpen
  \bibfield  {author} {\bibinfo {author} {\bibfnamefont {D.}~\bibnamefont
  {Ayres}} \emph {et~al.},\ }\href@noop {} {\  (\bibinfo {year} {2002})},\
  \Eprint {http://arxiv.org/abs/hep-ex/0210005} {arXiv:hep-ex/0210005 [hep-ex]}
  \BibitemShut {NoStop}%
\bibitem [{\citenamefont {McDonald}(2001)}]{McDonald:2001mc}%
  \BibitemOpen
  \bibfield  {author} {\bibinfo {author} {\bibfnamefont {K.~T.}\ \bibnamefont
  {McDonald}},\ }\href@noop {} {\  (\bibinfo {year} {2001})},\ \Eprint
  {http://arxiv.org/abs/hep-ex/0111033} {arXiv:hep-ex/0111033 [hep-ex]}
  \BibitemShut {NoStop}%
\bibitem [{\citenamefont {Asano}\ and\ \citenamefont
  {Minakata}(2011)}]{Asano:2011nj}%
  \BibitemOpen
  \bibfield  {author} {\bibinfo {author} {\bibfnamefont {K.}~\bibnamefont
  {Asano}}\ and\ \bibinfo {author} {\bibfnamefont {H.}~\bibnamefont
  {Minakata}},\ }\href {\doibase 10.1007/JHEP06(2011)022} {\bibfield  {journal}
  {\bibinfo  {journal} {JHEP}\ }\textbf {\bibinfo {volume} {06}},\ \bibinfo
  {pages} {022} (\bibinfo {year} {2011})},\ \Eprint
  {http://arxiv.org/abs/1103.4387} {arXiv:1103.4387 [hep-ph]} \BibitemShut
  {NoStop}%
\end{thebibliography}%

\end{document}